\documentclass[
    10pt,
    floatfix,
    twocolumn,
    groupedaddress,
    prb,
    a4paper,
    showpacs
]{revtex4}

\usepackage{graphicx}
\usepackage{amssymb}
\usepackage{amsmath}
\usepackage{color}

\newcommand{\intd}{d}
\newcommand{\intR}{\int_{-\infty}^\infty}

\newcommand{\imag}{\Im \textrm m}
\newcommand{\etal}{\mbox{\textit{et al.}}}
\newcommand{\siesta}{\textsc{Siesta}}
\newcommand{\transiesta}{\textsc{Tran\-siesta}}

\newcommand{\mysum}[2]{\sum\limits_{#1}^{#2}}
\newcommand{\myint}[2]{\int\limits_{#1}^{#2}}

\newcommand{\mybpar}[1]{\left( #1 \right)}
\newcommand{\beq}{\begin{equation}}
\newcommand{\eeq}{\end{equation}}
\newcommand{\beqa}{\begin{eqnarray}}
\newcommand{\eeqa}{\end{eqnarray}}

\newcommand{\Eqref}[1]{Eq.~(\ref{#1})}
\newcommand{\Figref}[1]{Fig.~\ref{#1}}

\newcommand{\Secref}[1]{Sec.~\ref{#1}}
\newcommand{\Appref}[1]{App.~\ref{#1}}

\begin{document}
\title{Inelastic transport theory from first-principles: \\methodology and applications
for nanoscale devices}
\date{\today}

%%%%%%%%%%%%%%%%%%%%%%%%%%%%%%%%%%%%%%%%%%%%%%%%%%%%%%%%%%%%%%%%%%%%%%%%%%%%%

\author{Thomas \surname{Frederiksen}}
\email{thf@mic.dtu.dk}
\author{Magnus \surname{Paulsson}}
\author{Mads \surname{Brandbyge}}
\author{Antti--Pekka \surname{Jauho}}
\affiliation{MIC -- Department of Micro and Nanotechnology, NanoDTU,
Technical University of Denmark,\\ {\O}rsteds Plads, Bldg.~345E,
DK-2800 Lyngby, Denmark}
\date{\today}

%%%%%%%%%%%%%%%%%%%%%%%%%%%%%%%%%%%%%%%%%%%%%%%%%%%%%%%%%%%%%%%%%%%%%%%%%%%%%%

\begin{abstract}
  We describe a first-principles method for calculating electronic structure,
  vibrational modes and frequencies, electron-phonon couplings, and inelastic
  electron transport properties of an atomic-scale device bridging two
  metallic contacts under nonequilibrium conditions. The method
  extends the density-functional codes {\siesta} and {\transiesta} that use atomic
  basis sets. The inelastic conductance characteristics are calculated using the
  nonequilibrium Green's function formalism, and the electron-phonon
  interaction is addressed with perturbation theory up to the level
  of the self-consistent Born approximation. While these calculations often are
  computationally demanding, we show how they can be approximated by a simple
  and efficient lowest order expansion.
  Our method also addresses effects of energy dissipation and local heating
  of the junction via detailed calculations of the power flow.
  We demonstrate the developed procedures by considering inelastic transport
  through atomic gold wires of various lengths, thereby extending the results
  presented in [Frederiksen \etal~Phys.~Rev.~Lett.~\textbf{93}, 256601
  (2004)]. To illustrate that the method applies more generally to molecular
  devices, we also calculate the inelastic current through different hydrocarbon
  molecules between gold electrodes. Both for the wires and the molecules our theory is in
  quantitative agreement with experiments, and characterizes the system-specific
  mode selectivity and local heating.
\end{abstract}

\maketitle

%%%%%%%%%%%%%%%%%%%%%%%%%%%%%%%%%%%%%%%%%%%%%%%%%%%%%%%%%%%%%%%%%%%%%%%%%%%%%%%%%%%%%
%\input{Introduction}
%%%%%%%%%%%%%%%%%%%%%%%%%%%%%%%%%%%%%%%%%%%%%%%%%%%%%%%%%%%%%%%%%%%%%%%%%%%%%%
\section{Introduction}
Electron transport in atomic-scale devices is an important research
area where both fundamental physics and technological opportunities
are simultaneously addressed.\cite{Cuniberti2005} Examples of novel
structures include molecules in self-assembled monolayers
(SAM),\cite{ReZhMu.97.Conductanceofmolecular} carbon nanotube based
components,\cite{AnLe.06.Physicsofcarbon}
nanowires,\cite{AgYeva.03.Quantumpropertiesof} and single-molecule
junctions.\cite{PaPaGo.02.Coulombblockadeand,
XuXiTa.03.Measurementsofsingle-molecule,
KuDaHj.03.Single-electrontransistorof,
XiXuTa.04.Measurementofsingle,MaWe.04.Statisticalanalysisof} Also
conventional lithography-based semiconductor electronics is rapidly
being pushed towards the scale where atomic features become
important. For example, the transistor gate oxide is now only a few
atomic layers thick.\cite{Roadmap2005}

The interaction between electrons and nuclear vibrations plays an
important role for the electron transport at the nanometer
scale,\cite{Kr.06.Electron-phononcouplingat,HoBoNe.06.transferofenergy}
and is being addressed experimentally in ultimate atomic-sized
systems.\cite{StReHo.98.Single-moleculevibrationalspectroscopy,
PaPaLi.00.Nanomechanicaloscillationsin,
AgUnRu.02.Onsetofenergy,SmNoUn.02.Measurementofconductance,LeLeKo.04.Electricalgenerationand,
DjThUn.05.Stretchingdependenceof} Effects on the electronic current
due to energy dissipation from electron-phonon (e-ph) interactions
are relevant, not only because they affect device characteristics,
induce chemical
reactions,\cite{PaLoSo.03.Selectivityinvibrationally} and ultimately
control the stability; these may also be used for spectroscopy to
deduce structural information---such as the bonding configuration in
a nanoscale junction---which is typically not accessible by other
techniques simultaneously with transport measurements.

% Inelastic effects in transport
The signatures of e-ph interaction have been observed in a variety
of nanosystems. In the late 1990s inelastic electron tunneling
spectroscopy (IETS) on single molecules was successfully
demonstrated using a scanning tunneling microscope
(STM).\cite{StReHo.98.Single-moleculevibrationalspectroscopy} Later,
in the quantum dot regime, measurements on a single C$_{60}$
transistor showed features indicating a strong coupling between
center-of-mass motion of the molecule and single-electron
hopping.\cite{PaPaLi.00.Nanomechanicaloscillationsin} Point contact
spectroscopy has also revealed phonon signals in the
high-conductance regime, e.g., in atomic
wires\cite{AgUnRu.02.Onsetofenergy} and individual
molecules.\cite{SmNoUn.02.Measurementofconductance} Most recently,
inelastic measurements have also been reported on SAMs of alkyl- and
$\pi$-conjugated molecular
wires.\cite{KuLaPa.04.Vibroniccontributionsto,WaLeKr.04.Inelasticelectrontunneling,Long}
These developments show the need for fully atomistic
\textit{quantitative} theories to accurately model structural,
vibrational, and transport properties of nanoscale systems.

% Quantum transport and DFT
The density functional theory (DFT) approach offers an atomistic
description of total energy properties of nanosystems without system
specific adjustable parameters. Furthermore, in combination with the
nonequilibrium Green's function (NEGF)
method\cite{Datta1995,Haug1996} it has recently become a popular
approach to quantum transport in atomic
structures.\cite{TaGuWa.01.Abinitiomodelinga,
BrMoOr.02.Density-functionalmethodnonequilibrium,
PaPeLo.02.First-principlesapproachto,XuDaRa.02.First-principlesbasedmatrix,
PeCa.04.Atomistictheoryof,RoGaBa.05.Towardsmolecularspintronics,
KeBaYa.05.Contactatomicstructure,
ThJa.05.Moleculartransportcalculations} From the comparison with
experimental data it has been established that total energy
properties such as atomic structure and vibrations in general are
well described by DFT with the local or gradient approximations for
exchange and correlation.\cite{BadeDa.01.Phononsandrelated} However,
while transport properties may also be calculated from DFT this is
not rigorously
justified.\cite{EvWeKo.04.Conductanceofmolecular,Evers} On the other
hand such an approach can serve as a good starting point for more
sophisticated approaches correcting for errors in, e.g., the
excitation spectrum, such as time-dependent
DFT,\cite{KuStAl.05.Time-dependentquantumtransport:} the GW
approximation,\cite{OnReRu.02.Electronicexcitations:density-functional,Thygesen,Darancet}
or self-interaction corrected
DFT.\cite{ToFiSa.05.Self-interactionerrorsin,Pemmaraju} In weakly
coupled molecular conductors electron-electron interaction effects
play a significant role. While some Coulomb blockade effects have
been described using spin-density functional
theory,\cite{Pa.05.Coulombblockadein} the correlation effects are
more complicated to treat. In this direction the addition of a
Hubbard-like term on top of the DFT Hamiltonian has been
used.\cite{FeCaDi.05.First-principlestheoryof} These more advanced
developments often come at the price of limitations to the size of
the systems that feasibly can be handled. It is therefore
interesting to investigate to what extent the conventional DFT-NEGF
can be used to model various transport properties.

% CONTENTS OF THIS PAPER
In this paper we present a scheme for including the effects of e-ph
interaction into one such DFT-NEGF method for electronic transport.
Specifically, we describe in detail our implementation of methods
based on a combination of the
{\siesta}\cite{SoArGa.02.SIESTAmethodab} and the
{\transiesta}\cite{BrMoOr.02.Density-functionalmethodnonequilibrium}
DFT computer codes. {\siesta} provides the fundamental
implementation of Kohn-Sham DFT in an atomic basis set for systems
described in a supercell representation (periodic boundary
conditions). {\transiesta}, on the other hand, uses the {\siesta}
framework to solve self-consistently the Kohn-Sham DFT equations for
the nonequilibrium electron density in the presence of a current
flow, taking into account the full atomistic structure of both
device and electrodes (no periodicity in the transport direction).
We describe how the {\siesta} and {\transiesta} methods have been
extended for inelastic transport analysis, which involves the
calculation of (i) relaxed geometries, (ii) vibrational frequencies,
(iii) e-ph couplings, and (iv) inelastic current-voltage
characteristics up to the level of the self-consistent Born
approximation (SCBA). We also describe approximations leading to a
lowest order expansion (LOE) of the SCBA expressions, which vastly
simplifies the computational
burden.\cite{PaFrBr.05.Modelinginelasticphonona,ViCuPa.05.Electron-vibrationinteractionin}

% OTHER'S WORK ON INELASTIC TRANSPORT
While there have already been many studies devoted to transport with
e-ph interaction based on model Hamiltonians emphasizing various
aspects of the
transport,\cite{NeShFi.01.Coherentelectron-phononcoupling,
MoHoTo.03.Inelasticcurrent-voltagespectroscopy,
BrFl.03.Vibrationalsidebandsand,MoTo.03.Electron-phononinteractionin,MiAlMi.04.Phononeffectsin,
HoBoFi.04.Powerdissipationin,
As.04.TheoryofInelastic,GaRaNi.04.Inelasticelectrontunneling,
NeFi.05.Vibrationalinelasticscattering,
YaWaWa.05.Electronictransportin,ViCuPa.05.Electron-vibrationinteractionin,
RyKe.05.Inelasticresonanttunneling,RyHaCu.06.Nonequilibriummolecularvibrons,
deMaAg.06.Universalfeaturesof,
GaNiRa.06.Resonantinelastictunneling,Mi.06.Anharmonicphononflow}
there has only been a handful based on a complete first-principles
description of all aspects of the e-ph transport problem (described
below). By this distinction we intend to emphasize approaches where
structural, vibrational, and transport properties are derived from
the knowledge of the elemental constituents only, i.e., without any
system-dependent adjustable parameters. So far these have almost
entirely been based on DFT for the electronic structure.

In the tunneling regime the atomic resolution of the STM has been
used to investigate spatial variations of the inelastic tunneling
process through adsorbed molecules on metallic surfaces.
Corresponding inelastic STM images were simulated theoretically by
Lorente and Persson with DFT and the Tersoff-Hamann
approach.\cite{LoPe.00.Theoreticalaspectsof,LoPeLa.01.Symmetryselectionrules}
Also controlled conformational changes, molecular motion, and
surface chemistry induced by the inelastic tunnel current in STM
have been
addressed.\cite{Ue.03.Motionsandreactions,UeMiLo.05.Adsorbatemotionsinduced,
LoRuTa.05.Single-moleculemanipulationand}

More recently the regime where an atomic-scale conductor is more
strongly coupled to both electrodes has also been investigated.
Based on a self-consistent tight-binding procedure with parameters
obtained from DFT,\cite{PeCa.04.Atomistictheoryof} Pecchia {\etal}
considered vibrational effects in octanethiols bonded to gold
electrodes using NEGF and the Born approximation (BA) for the e-ph
interaction.\cite{PeDiGa.04.Incoherentelectron-phononscattering}
Solomon {\etal} further used this method to simulate the
experimental IETS spectra of Wang
{\etal}\cite{WaLeKr.04.Inelasticelectrontunneling,
SoGaPe.06.Understandinginelasticelectron-tunneling} Sergueev {\etal}
studied a 1,4-benzenedithiolate molecule contacted by two aluminum
leads.\cite{SeRoGu.05.Abinitioanalysis} This study addressed the
bias dependence of the vibrational modes and e-ph couplings, but not
the inelastic current itself. While the vibrational spectrum was
found to be almost unchanged, a significant change in the e-ph
couplings was found at high bias voltages ($V_\textrm{bias}>$ 0.5
V). Chen {\etal} studied inelastic scattering and local heating in
an atomic gold contact, a thiol-bonded benzene, and
alkanethiols.\cite{ChZwDi.03.Localheatingin,
ChZwDi.04.Inelasticcurrent-voltagecharacteristics,
ChZwDi.05.Inelasticeffectson} The inelastic signals were calculated
using a golden-rule-type of expression and the DFT scattering states
where calculated using jellium
electrodes.\cite{DiLa.02.Transportinnanoscale} However, contrary to
experiments and most calculations on molecules---for example
Refs.\onlinecite{KuLaPa.04.Vibroniccontributionsto,WaLeKr.04.Inelasticelectrontunneling,
PeDiGa.04.Incoherentelectron-phononscattering,SoGaPe.06.Understandinginelasticelectron-tunneling,
JiKuLu.05.First-principlessimulationsof,TrRa.05.Modelinginelasticelectron,
PaFrBr.06.InelasticTransportthrough}---they predict conductance
decreases by the phonons for alkanethiols. Jiang {\etal} used a
related golden-rule approach for molecular
systems.\cite{JiKuLu.05.First-principlessimulationsof} Troisi
{\etal} suggested a simplified approach from which IETS signals can
be calculated approximately based on ab initio calculations for an
isolated cluster and neglecting the
electrodes.\cite{TrRaNi.03.Vibroniceffectsin,TrRa.05.Modelinginelasticelectron}
This scheme was shown to be suitable for the off-resonance regime,
i.e., when the molecular levels are far away from the Fermi level.
Their results compare well with experiments by Kushmerick
{\etal}\cite{KuLaPa.04.Vibroniccontributionsto} During the
development of the scheme presented here, we studied the same
molecular systems with similar
results.\cite{PaFrBr.05.Modelinginelasticphonona,PaFrBr.06.InelasticTransportthrough}
We also used it to model inelastic effects that can be observed in
atomic gold wires.\cite{FrBrLo.04.InelasticScatteringand}

\begin{figure}[t!]
\begin{center}
\includegraphics[width=\columnwidth]{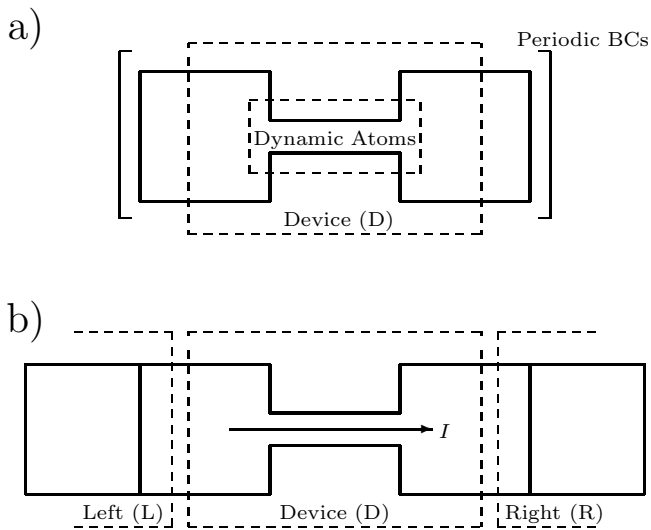}
\caption{Schematic of two generic system setups. (a) To calculate
vibrational frequencies and e-ph couplings with {\siesta} we use a
supercell setup with periodic boundary conditions (BCs) in all
directions. The cell contains the device region $D$ and possibly
some additional atom layers to come closer to a representation of
bulk electrodes. The dynamic atoms are a relevant subset of the
device atoms for which we determine the vibrations. (b) In the
transport setup we apply the {\transiesta} scheme where the central
region $D$ is coupled to fully atomistic semi-infinite electrodes
via self-energies, thereby removing periodicity along the transport
direction (the periodic BCs are retained in the transverse plane).}
\label{fig:TwoGenericUnitCells}
\end{center}
\end{figure}

\begin{figure*}[t!]
\begin{center}
\includegraphics[width=\textwidth]{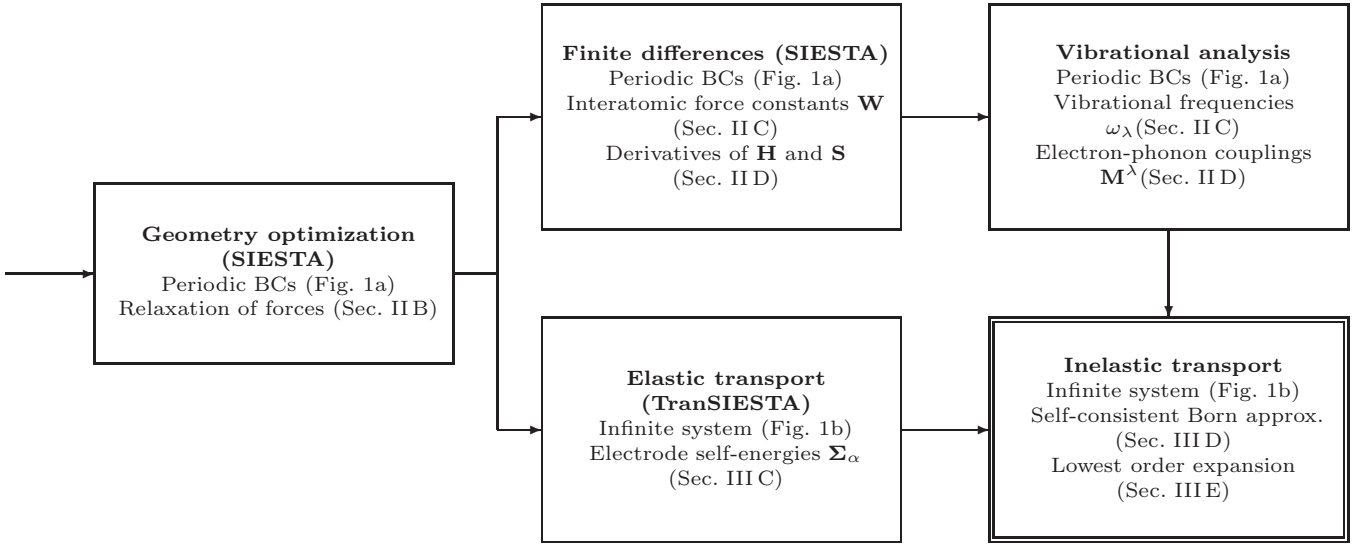}
\caption{Flow diagram for the complete analysis of the inelastic
transport properties of an atomic structure.}
\label{fig:FlowDiagram}
\end{center}
\end{figure*}

% OUTLINE
The paper is organized as follows. In
\Secref{sec:ElectronicStructureMethods} we communicate our
first-principles approach to obtain a Hamiltonian description of a
vibrating atomic-scale device bridging two metallic contacts, such
as schematically shown in \Figref{fig:TwoGenericUnitCells}.
Specifically we describe the use of {\siesta} to calculate
vibrational modes and e-ph couplings. Section~\ref{sec:Transport}
addresses the NEGF formalism used to calculate the inelastic
electron transport in steady state as well as the SCBA and LOE
schemes for the e-ph interaction. Electrode self-energies are
obtained using the {\transiesta} scheme. We further discuss local
heating effects and how various broadening mechanisms of the
inelastic signal can be addressed. The main steps of the method
presented in \Secref{sec:ElectronicStructureMethods} and
\ref{sec:Transport}, and how these depend on each other, are
schematically clarified in \Figref{fig:FlowDiagram}. In
\Secref{sec:AtomicGoldWires} and \ref{sec:HydrocarbonMolecules} we
illustrate our approach by corroborating and extending our previous
studies of atomic gold wires and hydrocarbon molecules.
Section~\ref{sec:AtomicGoldWires} gives results for an extensive set
of calculations for atomic gold wires of varying length and strain
conditions. From these calculations we identify a number of physical
effects, e.g., the evolution of a vibrational selection rule that
becomes more pronounced the longer the wire is.
Section~\ref{sec:HydrocarbonMolecules} illustrates that our method
is applicable to a wide range of systems, here exemplified by
different hydrocarbon molecules between gold surfaces. Both
applications also underline the usefulness of the LOE scheme, which
we validate by a comparison the full SCBA calculation. Finally in
\Secref{sec:Conclusions} we provide a summary of the paper and an
outlook.

%%%%%%%%%%%%%%%%%%%%%%%%%%%%%%%%%%%%%%%%%%%%%%%%%%%%%%%%%%%%%%%%%%%%%%%%%%%%%%%%%%%%%
%\input{ElectronicStructureMethods}
%%%%%%%%%%%%%%%%%%%%%%%%%%%%%%%%%%%%%%%%%%%%%%%%%%%%%%%%%%%%%%%%%%%%%%%%%%%%%%%%%%%%%
%%%%%%%%%%%%%%%%%%%%%%%%%%%%%%%%%%%%%%%%%%%%%%%%%%%%%%%%%%%%%%%%%%%%%%%%%%%%%%
\section{Electronic structure methods}
\label{sec:ElectronicStructureMethods} In this section we describe
our first-principles method to obtain a Hamiltonian description of a
vibrating atomic-scale device bridging to two metallic contacts. The
framework is the Density Functional Theory (DFT) and its numerical
implementation in the computer code
{\siesta}.\cite{SoArGa.02.SIESTAmethodab}

%%%%%%%%%%%%%%%%%%%%%%%%%%%%%%%%%%%%%%%%%%%%%%%%%%%%%%%%%%%%%%%%%%%%%%%%%%%%%%
\subsection{Vibrational Hamiltonian}
\label{sec:VibHam} The physical situations which we typically want
to describe can schematically be represented as a central device
region $D$ which is coupled to semi-infinite electrodes to the left
($L$) and right ($R$). This generic setup is shown in
\Figref{fig:TwoGenericUnitCells}(b).

We assume that the whole system under consideration can be described
by the following Hamiltonian
\begin{subequations}
\label{eq:VibHam}
\begin{eqnarray}
\widehat{H} &=& \widehat{H}^0_\textrm{e}+ \widehat{H}^0_\textrm{ph}+\widehat{H}_\textrm{e-ph}, \\
\widehat{H}^0_\textrm{e} &=& \sum_{i,j}H^0_{ij}{\widehat
c}^\dagger_{i}{\widehat c}^{\phantom\dagger}_{j},
\label{eq:MFHamiltonian}\\
\widehat{H}^0_\textrm{ph} &=& \sum_{\lambda}\hbar \omega_\lambda
{\widehat b}^\dagger_{\lambda}{\widehat
b}^{\phantom\dagger}_{\lambda},
\label{eq:HamFreeOscillators}\\
\widehat{H}_\textrm{e-ph} &=& \sum_{\lambda}\sum_{i,j}
M^{\lambda}_{ij}{\widehat c}^\dagger_{i}{\widehat
c}^{\phantom\dagger}_{j} ({\widehat b}^\dagger_{\lambda}+{\widehat
b}^{\phantom\dagger}_{\lambda}),\label{eq:ephCouplingHamiltonian}
\end{eqnarray}
\end{subequations}
where ${\widehat c}^\dagger_{i}$ and ${\widehat
b}^\dagger_{\lambda}$ are the electron and phonon creation
operators, respectively. Here $\widehat{H}^0_\textrm{e}$ is the
single-particle mean-field Hamiltonian describing electrons moving
in a static arrangement of the atomic nuclei,
$\widehat{H}^0_\textrm{ph}$ is the Hamiltonian of free uncoupled
phonons (oscillators), and $\widehat{H}_\textrm{e-ph}$ is the e-ph
coupling within the harmonic approximation. For simplicity, we
present in this paper a formulation for spin-independent problems.
The generalization to include spin-polarization is straightforward.

The Hamiltonian \Eqref{eq:VibHam} naturally arises from the
adiabatic approximation of Born-Oppenheimer in which the timescales
of electronic and vibrational dynamics are
separated.\cite{HoBoNe.06.transferofenergy} Since the electrons move
on a much shorter timescale than the heavy nuclei, the adiabatic
approximation states that the electronic Hamiltonian depends
parametrically on the nuclear coordinates, i.e., that
$\widehat{H}_\textrm{e}=\widehat{H}_\textrm{e}(\textbf{Q})$ where
$\textbf{Q}\equiv\textbf{R}-\textbf{R}^0$ is a displacement variable
around the equilibrium configuration $\textbf{R}^0$. Next, limiting
ourselves to small displacements we can expand the electronic
Hamiltonian to lowest order in $\textbf{Q}$
\begin{eqnarray}
\label{eq:BO-expansion} \widehat{H}_\textrm{e}&\approx&
\widehat{H}^0_\textrm{e}
    +\sum_{I\nu}\frac{\partial \widehat{H}_\textrm{e}}{\partial Q_{I\nu}}\Big|_{Q=0}
    Q_{I\nu},
\end{eqnarray}
where index $I$ runs over all dynamic nuclei and $\nu=x,y,z$ over
spatial directions. Imposing a transformation into normal mode
coordinates (and the usual canonical quantization of position and
momentum operators) we can rewrite \Eqref{eq:BO-expansion} into
\begin{eqnarray}
\label{eq:HamExpansionNormCoord}
 \widehat{H}_\textrm{e} &\approx&
\widehat{H}^0_\textrm{e}
    +\sum_{I\nu} \frac{\partial \widehat{H}_\textrm{e}}{\partial Q_{I\nu}}\Big|_{Q=0}
        \sum_\lambda \textrm{v}^{\lambda}_{I\nu} \sqrt{\frac{\hbar}{2M_I\omega_\lambda}}
        ({\widehat b}^\dagger_{\lambda}+{\widehat b}^{\phantom\dagger}_{\lambda}),\nonumber\\
\end{eqnarray}
where $M_I$ is the mass of ion $I$ and
$\textbf{\textrm{v}}^\lambda=\{\textrm{v}_{I\nu}^\lambda\}$ is the
ionic displacement vector of normal mode $\lambda$ with frequency
$\omega_\lambda$ normalized according to
$\textbf{v}^{\lambda}\cdot\textbf{v}^{\lambda} = 1$. From
\Eqref{eq:HamExpansionNormCoord} we identify the e-ph coupling
matrix elements of \Eqref{eq:ephCouplingHamiltonian} as
\begin{eqnarray}
\label{eq:ephCouplingMatrixElements} M_{ij}^{\lambda} &=&
\sum_{I\nu} \langle i | \frac{\partial
\widehat{H}_\textrm{e}}{\partial
Q_{I\nu}}|j\rangle\raisebox{-4pt}{${}_{Q=0}$}\,
\textrm{v}^{\lambda}_{I\nu} \sqrt{\frac{\hbar}{2M_I\omega_\lambda}}.
\end{eqnarray}
In the following sections we describe how we determine the detailed
geometry, the vibrational modes, and the e-ph couplings from DFT.

%%%%%%%%%%%%%%%%%%%%%%%%%%%%%%%%%%%%%%%%%%%%%%%%%%%%%%%%%%%%%%%%%%%%%%%%%%%%%%
\subsection{SIESTA approach and geometry optimization}
In our numerical approach we use the {\siesta} implementation of
DFT.\cite{SoArGa.02.SIESTAmethodab} This code treats exchange and
correlation within the local density approximation (LDA) or the
generalized gradient approximation (GGA). The core electrons are
described with pseudopotentials.

The main reason why {\siesta} is particularly suitable starting
point for transport calculations is that the valence electrons are
described in a localized basis set that allows for an unambiguous
partitioning of the system into leads and device,
cf.~\Figref{fig:TwoGenericUnitCells}(b), thereby making it possible
to calculate the flux of electrons (the necessity of this
partitioning for transport calculations is discussed further in
\Secref{sec:Transport}). The basis orbitals $\{|i\rangle\}$ are
strictly localized approximations to atomic orbitals with a given
cutoff radius and centered at the positions of the nuclei of the
structure. Importantly, this local electronic basis is nonorthogonal
with overlap matrix elements $S_{ij}=\langle i|j\rangle$.

In this tight-binding like basis we use the Kohn-Sham Hamiltonian
from {\siesta} as the mean-field Hamiltonian in
\Eqref{eq:MFHamiltonian}. We initially construct a periodic
supercell [\Figref{fig:TwoGenericUnitCells}(a)], and use it as an
approximation to the full transport setup
[\Figref{fig:TwoGenericUnitCells}(b)] for relaxing the device atoms,
and to obtain vibrational frequencies and e-ph couplings. We note
that this step leads to a determination of the quantities in
equilibrium. In principle, these could also be calculated under
nonequilibrium conditions by retaining the full transport structure
of \Figref{fig:TwoGenericUnitCells}(b). Recently, Sergueev {\etal}
showed this to be important for relatively high voltages
($eV\gg\hbar\omega_\lambda$).\cite{SeRoGu.05.Abinitioanalysis}
However, for the low-bias regime considered in this paper the
equilibrium calculation is sufficient.

A fairly accurate relaxation is an important prerequisite for the
subsequent calculation of vibrational modes. The atoms in the device
region are therefore typically relaxed until the forces acting on
the dynamic atoms all are smaller than
$F_{I\nu}(\textbf{R}^0)<F_\textrm{max}=0.02$ eV/{\AA}. Compared with
other error sources in the calculations little is gained by lowering
this criteria. \label{sec:GeomOpt}

%%%%%%%%%%%%%%%%%%%%%%%%%%%%%%%%%%%%%%%%%%%%%%%%%%%%%%%%%%%%%%%%%%%%%%%%%%%%%%
\subsection{Vibrational modes}
\label{sec:VibModes} The starting point for our description of the
nuclear vibrations is the Born-Oppenheimer total energy surface
$E(\mathbf R)$ (BOS) and its derivatives with respect to the nuclear
coordinates. For a thorough review on phonons from DFT we refer the
reader to the paper by Baroni
\etal\cite{BadeDa.01.Phononsandrelated} From the BOS we define the
matrix of interatomic force constants (usually called the Hessian or
dynamic matrix) as
\begin{eqnarray}
\label{eq:DefHessian} C_{I\nu;J\mu} &\equiv& \frac{\partial^2
E(\textbf{R})}{\partial R_{I\nu}\partial
R_{J\mu}}\Big|_{\textbf{R}=\textbf{R}^0},
\end{eqnarray}
where $\textbf{R}\equiv\{\textbf{R}_I\}$ denotes the full set of
nuclear coordinates, and $\textbf{R}_I\equiv\{R_{I\nu}\}$ the
coordinates of nucleus $I$ with mass $M_I$ (not to be confused with
the e-ph coupling elements $M^\lambda_{ij}$). Within the harmonic
approximation we can write the time-dependent displacement variable
as
\begin{eqnarray}
\label{eq:TDDisplVar} \textbf{Q}_I(t) &=&
\textbf{R}_I(t)-\textbf{R}_I^0 \equiv \textbf{Q}_I e^{i\omega t}.
\end{eqnarray}
Inserting \Eqref{eq:DefHessian} and (\ref{eq:TDDisplVar}) into
Newton's second law of motion
\begin{eqnarray}
M_I \frac{\partial^2 \textbf{R}_I}{\partial t^2}=
\textbf{F}_I(\textbf{R}) = -\frac{\partial E(\textbf{R})}{\partial
\textbf{R}_I},
\end{eqnarray}
we have
\begin{eqnarray}
\label{eq:eigenproblem} -\omega^2 M_I Q_{I\nu} &=&
-\sum_{J\mu}C_{I\nu;J\mu} \,Q_{J\mu}.
\end{eqnarray}
Introducing boldface notation also for matrices we can rewrite
\Eqref{eq:eigenproblem} to the following ordinary eigenvalue problem
\begin{eqnarray}
\label{eq:eigenproblemMatrixform}
(\omega^2\textbf{1}-\textbf{W})\textbf{v} = 0,
\end{eqnarray}
where the mass-scaled matrix of interatomic force constants is
\begin{eqnarray}
W_{I\nu,J\mu}\equiv \frac {C_{I\nu;J\mu}}{\sqrt{M_IM_J}},
\end{eqnarray}
and $\textbf{v}_I=\sqrt{M_I}\textbf{Q}_I$. Thus, the vibrational
frequency $\omega_\lambda$ and mode
$\textbf{v}^{\lambda}=\{\textbf{v}^{\lambda}_I\}$ belong to the
eigensolution $(\omega_\lambda^2,\textbf{v}^{\lambda})$ to
\Eqref{eq:eigenproblemMatrixform} where we normalize the vectors as
$\textbf{v}^{\lambda}\cdot\textbf{v}^{\lambda} = 1$.

\begin{figure}[t]
  \centering
  \includegraphics[width=.8\columnwidth]{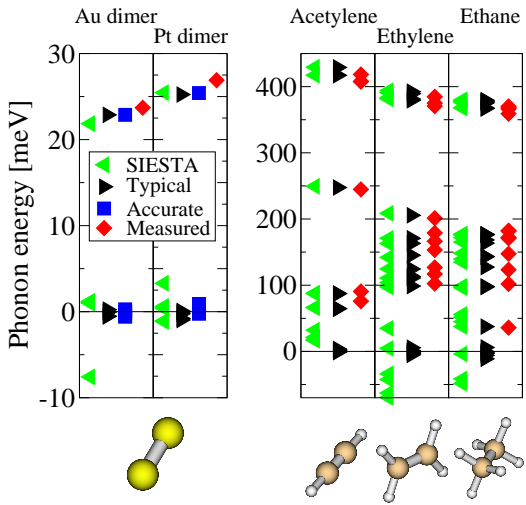}
  \caption{(Color online) Vibrational frequencies calculated for some simple molecules (Au${}_2$
  and Pt${}_2$, acetylene $\textrm{C}_2\textrm{H}_2$,
  ethylene $\textrm{C}_2\textrm{H}_4$, and
  ethane $\textrm{C}_2\textrm{H}_6$). The results obtained directly from {\siesta} are
  shown together with those of our scheme (typical/accurate) based on the correction
  \Eqref{eq:CorrectedFCmatrix}. The different calculational settings are described in the text. For comparison
  the experimentally measured values of the frequencies are also given.\cite{HeKa.00.Relativisticall-electroncoupled-cluster,
  AiMo.02.Rotationallyresolvedspectroscopy,NISTChemWebbook}
  To indicate the accuracy of the calculations the numerical values for the zero-frequency modes
  (translation/rotation) are included, where negative values correspond to imaginary frequencies.}
  \label{fig:Comparison_SimpleMolecules}
\end{figure}

\begin{figure}[t]
  \centering
  \includegraphics[width=.9\columnwidth]{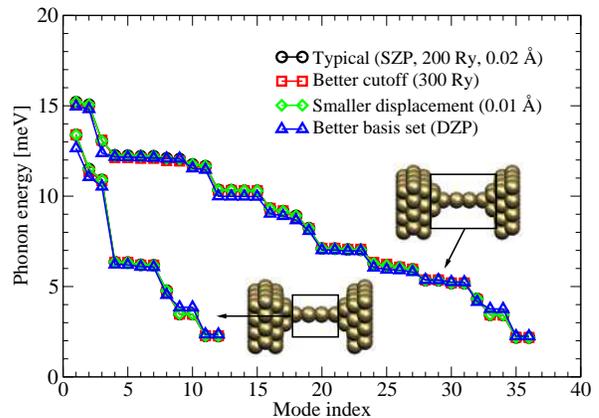}
  \caption{(Color online) Convergence of calculated vibrational frequencies for a 4-atom Au wire
  with the most important DFT settings. For each of the two choices for the vibrational
  region (as indicated with boxes) the reference calculation---carried out with
  SZP, a 200 Ry real space grid energy cutoff,
  and 0.02 {\AA} finite displacements---and other three separate calculations (with one
  of the settings improved at a time) yield essentially the same results for the phonon
  energy $\hbar\omega_\lambda$ versus mode index $\lambda$.}
  \label{fig:CalcSettings}
\end{figure}

Atomic forces $\mathbf{F}_I =\{F_{I\nu}\}$ are directly obtained by
{\siesta} along with the total energy
calculation.\cite{SoArGa.02.SIESTAmethodab} This allows us to
approximate the dynamic matrix by finite differences (``frozen
phonons''), either by
\begin{eqnarray}
\label{eq:HessianPlusMinus} \overline{C}^{(\pm)}_{I\nu;J\mu} &=&
-\frac{{F}_{I\nu}(\pm Q_{J\mu})-{F}_{I\nu}(0)}{\pm Q_{J\mu}},
\end{eqnarray}
or, numerically more accurately, by
\begin{eqnarray}
\label{eq:HessianDiff} \overline{C}_{I\nu;J\mu} &=&
-\frac{{F}_{I\nu}(Q_{J\mu})-{F}_{I\nu}(-Q_{J\mu})}{2Q_{J\mu}},
\end{eqnarray}
where the bar denotes the finite difference approximation. The
quantities in \Eqref{eq:HessianPlusMinus} and (\ref{eq:HessianDiff})
are thus readily determined. Typically we use a finite displacement
of the dynamic atoms in each spatial direction of $Q_{J\mu} = \pm
0.02$ {\AA}.

While the {\siesta} calculations for $\overline{C}_{I\nu;J\mu}$ are
generally straightforward, we have observed that {\siesta} has
difficulties in estimating the change in force on the atom
\textit{that is being displaced}. This problem relates to the
so-called egg-box effect, i.e., the movement of basis orbitals
(which follows the nuclear positions) with respect to the real space
integration grid.\cite{SoArGa.02.SIESTAmethodab} As a result,
phonons cannot be accurately obtained directly from
$\overline{C}_{I\nu;J\mu}$. To circumvent this technicality we
impose momentum conservation (in each direction $\nu$) via
$\sum_I\Delta F_{I\nu}=0$, which then determines the diagonal
elements according to
\begin{eqnarray}
\label{eq:CorrectedFCmatrix} \overline{\overline{C}}_{I\nu;J\mu}
&=&\left\{
  \begin{array}{cc}
    \overline{C}_{I\nu;J\mu}, & I\neq J \\
    -\sum_{K\neq I}\overline{C}_{I\nu;K\mu}, & I=J
  \end{array}\right.
\end{eqnarray}
where the $K$-sum runs over all atoms in the supercell. Finally,
since $\partial^2 E/\partial R_{I\nu}\partial R_{J\mu}=\partial^2
E/\partial R_{J\mu}\partial R_{I\nu}$ we apply a numerical
symmetrization of the force constants in the dynamic region. As a
check we always verify that the frequencies calculated from the
dynamic matrices with forward, backward, and combined displacements
[\Eqref{eq:HessianPlusMinus} and (\ref{eq:HessianDiff})] are roughly
the same, indicating that the harmonic approximation is not violated
with the given displacement amplitude $Q_{J\mu}$.

The eigenvalues $\{\omega_\lambda^2\}$ corresponding to the
symmetric matrix $\mathbf{W}$ are real numbers. Some of these may
however become negative leading to imaginary frequencies
$\{\omega_\lambda\}$, indicating that the atomic configuration
$\mathbf R^0$ is in fact not describing a true energy minimum of the
BOS. We shall denote such imaginary phonon frequencies by negative
values in \Figref{fig:Comparison_SimpleMolecules} and
\ref{fig:EtotVsLength}.

A comparison between calculated and experimentally measured
vibrational frequencies for some simple molecules is shown in
\Figref{fig:Comparison_SimpleMolecules}. Specifically we include
both the frequencies obtained directly with {\siesta} (from
$\overline{C}_{I\nu;J\mu}$) as well as those of our scheme based on
the correction \Eqref{eq:CorrectedFCmatrix}. In the calculations for
the dimers the important settings correspond to either a 200 Ry
cutoff for the real space grid integration and a single-zeta plus
polarization (SZP) basis set ({\siesta}/typical), or a 400 Ry cutoff
and a double-zeta plus polarization (DZP) basis set (accurate). For
the hydrocarbon molecules the settings are 200 Ry cutoff and DZP
basis set. In all calculations the displacement amplitude is
$Q_{J\mu} = 0.02$ {\AA}. The figure illustrates that our scheme
presented above leads to a quite accurate description of the
vibrational frequencies. We thus see no need to resort to a
frequency scaling which is sometimes invoked in DFT calculations.
Further, the figure shows that the use of momentum conservation for
correcting elements in the {\siesta} dynamic matrix improves the
calculation, in particular the determination of low frequency modes
(including the zero-frequency rotation/translation modes of isolated
molecules).

As an illustration of the convergence of the phonon energies with
respect to some important DFT settings for larger systems, we show
in \Figref{fig:CalcSettings} the calculated phonon energies for two
different sizes of the dynamic region of a four atom gold wire
(shown in the insets). We obtain almost identical frequencies by
increasing the real space integration grid cutoff from 200 Ry to 300
Ry, by using a DZP basis set instead of a SZP, or by changing the
finite displacements $Q_{J\mu}$ from 0.02 {\AA} to 0.01 {\AA}. We
expect the overall accuracy of these calculations to be
representative not only for isolated molecules but also for larger
periodic systems as well as systems involving other elements.

%%%%%%%%%%%%%%%%%%%%%%%%%%%%%%%%%%%%%%%%%%%%%%%%%%%%%%%%%%%%%%%%%%%%%%%%%%%%%%
\subsection{Electron-phonon couplings}
\label{sec:ephCouplings} In order to compute the e-ph coupling
matrices $\mathbf{M}^\lambda\equiv\{\{M^\lambda_{ij}\}\}$ we have
modified {\siesta} to output the Kohn-Sham Hamiltonian matrices
$\mathbf{H}(\mathbf{Q})\equiv\{\{\langle
i|\widehat{H}_\textrm{e}|j\rangle\}\}$ for each of the displaced
configurations.

The complicated part of the e-ph couplings in
\Eqref{eq:ephCouplingMatrixElements} is the evaluation of matrix
elements of gradients of the Hamiltonian operator. Rewriting this
part as
\begin{eqnarray}
\langle i|\frac{\partial \widehat{H}_\textrm{e}}{\partial
Q_{I\nu}}|j\rangle &=&
    \frac{\partial\langle i|\widehat{H}_\textrm{e}|j\rangle}{\partial Q_{I\nu}}
    -\langle i'|\widehat{H}_\textrm{e}|j\rangle -\langle i|\widehat{H}_\textrm{e}
    |j'\rangle,\qquad
\end{eqnarray}
where $|i'\rangle\equiv \partial|i\rangle/\partial Q_{I\nu}$
represents the change in basis orbitals with displacements, and
using the identity
\begin{eqnarray}
\sum_{ij}|i\rangle (\mathbf S^{-1})_{ij}\langle j| = 1,
\end{eqnarray}
where $\mathbf S \equiv\{\{\langle i|j\rangle\}\}$ is the overlap
matrix, we arrive at a form suitable for numerical evaluation

\begin{eqnarray}
\label{eq:DerivHamiltonianOperator} \langle i|\frac{\partial
\widehat{H}_\textrm{e}}{\partial Q_{I\nu}}|j\rangle &=&
\frac{\partial\langle i|\widehat{H}_\textrm{e}|j\rangle}{\partial
Q_{I\nu}}\nonumber\\
    &&-\sum_{kl}\langle i'|k\rangle (\mathbf S^{-1})_{kl}\langle l|\widehat{H}_\textrm{e}|j\rangle\nonumber\\
    &&-\sum_{kl}\langle i|\widehat{H}_\textrm{e} |k\rangle (\mathbf S^{-1})_{kl}\langle l|j'\rangle.
\end{eqnarray}
The first term on the right hand side in
\Eqref{eq:DerivHamiltonianOperator} can be approximated by finite
differences of Hamiltonian matrices. The factors $\langle
i'|k\rangle$ and $\langle l|j'\rangle$ are derivatives of the
orbital overlaps, which we readily determine from finite differences
via six separate runs that include both the original structure as
well as the whole structure displaced by $\pm Q_{J\mu}$ along each
spatial direction. We note that with the calculation of $\langle
i'|k\rangle$ and $\langle l|j'\rangle$ we avoid the further
approximations for the e-ph couplings proposed by Head--Gordon and
Tully\cite{HETU.92.VIBRATIONAL-RELAXATIONONMETAL-SURFACES} which we
have used previously.\cite{FrBrLo.04.InelasticScatteringand}

In some cases, if one works with a relatively small supercell, the
calculated Fermi energy may change slightly between the displaced
configurations of a given system. Since the real physical systems
are essentially infinite, such shifts in the Fermi energy are
artificial finite-size effects. To compensate for this we choose to
measure all energies with respect to the Fermi energy of the relaxed
structure $\varepsilon_\textrm{F}^0=\varepsilon_\textrm{F}(\mathbf
R^0)$, i.e., to shift the displaced Hamiltonians according to
\begin{eqnarray}
\overline{\mathbf H}(Q_{I\nu})&\equiv &{\mathbf H}(Q_{I\nu})
-[\varepsilon_\textrm{F}(Q_{I\nu})-\varepsilon_\textrm{F}^0]
\mathbf{S}(Q_{I\nu}).\nonumber\\
\end{eqnarray}
The finite difference approximation to the first term in
\Eqref{eq:DerivHamiltonianOperator}---the derivative of the
Hamiltonian matrix---may thus be written as
\begin{eqnarray}
\frac{\partial \overline{\mathbf H}}{\partial Q_{I\nu}}\Big|_{Q=0}
&\approx & \frac{1}{2Q_{I\nu}}\Big\{{\mathbf H}(Q_{I\nu})-{\mathbf
H}(-Q_{I\nu})\nonumber\\
&&\qquad\quad
-[\varepsilon_\textrm{F}(Q_{I\nu})-\varepsilon_\textrm{F}(-Q_{I\nu})]\mathbf
S^0\Big\},\nonumber\\
\end{eqnarray}
thereby completing the necessary steps to evaluate the e-ph coupling
matrix elements.

%%%%%%%%%%%%%%%%%%%%%%%%%%%%%%%%%%%%%%%%%%%%%%%%%%%%%%%%%%%%%%%%%%%%%%%%%%%%%%%%%%%%%
%\input{ElasticAndInelasticTransport}
%%%%%%%%%%%%%%%%%%%%%%%%%%%%%%%%%%%%%%%%%%%%%%%%%%%%%%%%%%%%%%%%%%%%%%%%%%%%%%%%%%%%%
%%%%%%%%%%%%%%%%%%%%%%%%%%%%%%%%%%%%%%%%%%%%%%%%%%%%%%%%%%%%%%%%%%%%%%%%%%%%%%
\section{Elastic and inelastic transport: The NEGF formalism}
\label{sec:Transport} In this section we describe how the NEGF
formalism is used to calculate the stationary electron transport
through a region in space with an e-ph interaction. The basic ideas
go back to the seminal work by Caroli
\etal\cite{CASACO.72.DIRECTCALCULATIONOF} but we shall use the later
formulation by Meir and
Wingreen.\cite{MEWI.92.LANDAUERFORMULACURRENT,JaWiMe.94.TIME-DEPENDENTTRANSPORTIN,Haug1996}

The starting point in the NEGF approach is a formal partitioning of
the system into a central device region (where interactions may
exist) and noninteracting leads.\footnote{Non-partitioning schemes
have also been proposed, e.g., by G.~Stefanucci and C.-O. Almbladh,
Phys.~Rev.~B \textbf{69}, 195318 (2004).} This partitioning was
sketched in \Figref{fig:TwoGenericUnitCells}(b). The e-ph
interaction is treated with diagrammatic perturbation theory. Below
we describe the SCBA as well as further approximations leading to
the computationally inexpensive LOE scheme. In addition, we discuss
local heating effects and how various broadening mechanisms of the
inelastic signal are addressed.

%%%%%%%%%%%%%%%%%%%%%%%%%%%%%%%%%%%%%%%%%%%%%%%%%%%%%%%%%%%%%%%%%%%%%%%%%%%%%%
\subsection{System partitioning}

The physical system of interest sketched in
\Figref{fig:TwoGenericUnitCells}(b) is infinite and
\emph{non}-periodic. For this setup let us initially consider the
electronic and vibronic problems separately and return later to the
treatment of their mutual interaction.

The use of a local basis in {\siesta} allows us to partition the
(bare) electronic Hamiltonian $\mathbf{H}\equiv\{\{H^0_{ij}\}\}$ and
overlap matrix $\mathbf{S}\equiv\{\{S_{ij}\}\}$ into
\begin{eqnarray}
\mathbf{H} &=& \left(
    \begin{array}{ccc}
    \mathbf{H}_{L} & \mathbf{H}_{LD} & 0 \\
    \mathbf{H}_{DL} & \mathbf{H}_{D} & \mathbf{H}_{DR} \\
    0 & \mathbf{H}_{RD} & \mathbf{H}_{R} \\
  \end{array}\right),\\
\mathbf{S} &=& \left(
    \begin{array}{ccc}
    \mathbf{S}_{L} & \mathbf{S}_{LD} & 0 \\
    \mathbf{S}_{DL} & \mathbf{S}_{D} & \mathbf{S}_{DR} \\
    0 & \mathbf{S}_{RD} & \mathbf{S}_{R} \\
  \end{array}\right),
\end{eqnarray}
in which the direct couplings and overlaps between leads $L$ and $R$
are strictly zero (provided that the central region is sufficiently
large).

In a similar fashion, since interatomic forces are short ranged, the
mass scaled dynamic matrix $\mathbf{W}$ [\Eqref{eq:DefHessian}] can
be partitioned into
\begin{eqnarray}
\mathbf{W} &=& \left(
    \begin{array}{ccc}
    \mathbf{W}_{L} & \mathbf{W}_{LD} & 0 \\
    \mathbf{W}_{DL} & \mathbf{W}_{D} & \mathbf{W}_{DR} \\
    0 & \mathbf{W}_{RD} & \mathbf{W}_{R} \\
  \end{array}\right),
\end{eqnarray}
where the direct coupling between leads $L$ and $R$ is neglected.

The infinite dimensionality of the electronic and vibrational
problem can effectively be addressed with the use of Green's
function techniques. For the electronic part we define the retarded
electronic single-particle Green's function
$\mathbf{G}^{0,r}(\varepsilon)$ as the inverse of
$[(\varepsilon+i\eta)\mathbf{S}-\mathbf{H}]$ where $\eta=0^+$. It is
then possible to write its representation in the device region $D$
as
\begin{eqnarray}
\mathbf{G}_D^{0,r}(\varepsilon) &=&
    [(\varepsilon +i\eta)\mathbf{S}_{D}-\mathbf{H}_{D}
    -\mathbf{\Sigma}_L^r(\varepsilon)-\mathbf{\Sigma}_R^r(\varepsilon)]^{-1},\qquad
\end{eqnarray}
where the self-energy due to the coupling to the left lead is
$\mathbf{\Sigma}^r_L(\varepsilon)= (\mathbf{H}_{DL}-\varepsilon
\mathbf{S}_{DL})\mathbf{g}_{L}^r(\varepsilon)
(\mathbf{H}_{LD}-\varepsilon \mathbf{S}_{LD})$ and similarly for the
right lead. Here, $\mathbf{g}_{\alpha}^r(\varepsilon)$ is the
retarded electronic ``surface'' Green's function of lead
$\alpha=L,R$ which can be calculated effectively for periodic
structures by recursive
techniques.\cite{SASARU.85.HIGHLYCONVERGENTSCHEMES} The quantities
$\mathbf{\Sigma}^r_\alpha(\varepsilon)$ are directly available from
{\transiesta}.\cite{BrMoOr.02.Density-functionalmethodnonequilibrium}
Note that Green's functions calculated without the e-ph interaction
are denoted with a superscript ``0''.

Similarly, for the vibrational part we can define the retarded
phonon Green's function $\mathbf{D}^{0,r}(\omega)$ as the inverse of
$[(\omega+i\eta)^2\mathbf{1}-\mathbf{W}]$, and write its
representation in the device region $D$ as
\begin{eqnarray}
\label{eq:PhononRetGreensFctWithLRselfenergies}
\mathbf{D}_D^{0,r}(\omega) &=&
    [(\omega +i\eta)^2\mathbf{1}-\mathbf{W}_{D}
    -\mathbf{\Pi}_L^r(\omega)-\mathbf{\Pi}_R^r(\omega)]^{-1},\nonumber\\
\end{eqnarray}
where the self-energies due to the coupling to the left and right
regions are $\mathbf{\Pi}^r_L(\omega)=
\mathbf{W}_{DL}\mathbf{d}_{L}^r(\omega)\mathbf{W}_{LD}$ and
$\mathbf{\Pi}^r_R(\omega)=
\mathbf{W}_{DR}\mathbf{d}_{R}^r(\omega)\mathbf{W}_{RD}$,
respectively. Here, $\mathbf{d}_{\alpha}^r(\omega)$ is the retarded
phonon ``surface'' Green's function which again can be calculated by
the recursion techniques mentioned above.

Note that the boldface matrix notation used for both electronic and
vibrational quantities refers to different vector spaces: Indices in
the electronic case refer to the basis orbitals and in the phonon
case to real space coordinates. In addition, the electronic problem
is treated directly in a nonorthogonal basis. The validity of the
nonorthogonal formulation has been discussed for the elastic
scattering problem in
Refs.~\onlinecite{BrKoTs.99.Conductionchannelsat,Zahid2003} and more
recently including interactions in
Ref.~\onlinecite{Th.06.Electrontransportthrough}.

Since we are interested in the interaction of the electronic current
with vibrations localized in the device region, we invoke the
\emph{ansatz} that---to a first approximation---we can disregard the
phonon lead self-energies $\mathbf{\Pi}_{\alpha}^r(\omega)$ and only
describe the device region by
\begin{eqnarray}
\mathbf{D}_D^{0,r}(\omega) &\approx&
[(\omega+i\eta)^2\mathbf{1}-\mathbf{W}_D]^{-1},
\label{eq:PhononGFapprox}
\end{eqnarray}
which in terms of the eigensolutions
$(\omega_\lambda^2,\mathbf{v}^\lambda)$ to
\Eqref{eq:eigenproblemMatrixform} can be written in a spectral
representation
\begin{eqnarray}
\mathbf{D}_D^{0,r}(\omega) &\approx& \sum_\lambda
\frac{\mathbf{v}^{\lambda} \otimes
\mathbf{v}^{\lambda}}{(\omega+i\eta)^2-\omega_\lambda^2}\nonumber\\
    &=& \sum_\lambda \mathbf{v}^{\lambda} \otimes
\mathbf{v}^{\lambda}\frac{d^r_0(\lambda,\omega)}{2\omega_\lambda},
\end{eqnarray}
where the free phonon Green's functions are\cite{Haug1996}
\begin{eqnarray}
d^{r,a}_0(\lambda,\omega)
    &=& \frac{1}{\omega-\omega_\lambda \pm i\eta}-\frac{1}{\omega+\omega_\lambda \pm i\eta},\\
\label{eq:92:PhGF0-lessergreater} d^\lessgtr_0(\lambda,\omega)
    &=&- 2\pi i [  \langle n_\lambda\rangle \delta(\omega\mp\omega_\lambda)\nonumber\\
        &&\qquad+(\langle n_\lambda\rangle + 1) \delta(\omega \pm \omega_\lambda) ],
\end{eqnarray}
with $\langle n_\lambda\rangle $ being the expectation value of the
occupation in mode $\lambda$. The lesser and greater Green's
functions stated above are used in \Secref{sec:SCBA} (transformed
into energy domain via $\omega\mapsto\hbar\omega$).

The validity of the approximation \Eqref{eq:PhononGFapprox} can be
investigated by calculating the correct phonon Green's function
according to \Eqref{eq:PhononRetGreensFctWithLRselfenergies}, and
then project the corresponding local density of states (per energy
via $\omega^2\mapsto\varepsilon$) onto each eigenmode
$\mathbf{v}^\lambda$ of the dynamic region (with fixed electrodes),
i.e., to determine
\begin{eqnarray}
\label{eq:PhononPDOS} B_\lambda(\varepsilon)\equiv -4 \varepsilon\,
\imag [ (\mathbf{v}^\lambda)^T \mathbf{D}_D^{0,r}(\varepsilon)
\mathbf{v}^\lambda],
\end{eqnarray}
satisfying the sum rule
\begin{eqnarray}
\label{eq:sumrule-B} \int_{0}^\infty \frac{\intd
\varepsilon}{2\pi}B_\lambda(\varepsilon) = 1.
\end{eqnarray}
If the mode $\mathbf{v}^\lambda$ is a true localized modes for the
extended system, then the projection $B_\lambda(\varepsilon)$
resembles a sharp resonance around the phonon energy
$\hbar\omega_\lambda$. In practice, $\{\mathbf{v}^\lambda\}$ are not
exact eigenmodes of the extended system, and the resonances hence
acquire finite widths. This broadening characterizes the damping
(within the harmonic approximation) of the modes by the coupling to
the electrodes. If the broadening is small compared with the phonon
energy (weak coupling to the bulk), then the projection can be
described by a Lorentzian
\begin{eqnarray}
\label{eq:LorentzBroadening} B_\lambda(\varepsilon)\approx \frac{2
\hbar\gamma_\textrm{damp}^\lambda}{(\varepsilon-\hbar\omega_\lambda)^2+
(\hbar\gamma_\textrm{damp}^\lambda)^2},
\end{eqnarray}
where $\hbar\gamma_\textrm{damp}^\lambda$ is the half width half max
(HWHM) value that transforms in time domain into an exponential
decay of the phonon population with an average lifetime
$\tau_\textrm{ph}^\lambda=1/\gamma_\textrm{damp}^\lambda$. We will
return to the question of a finite phonon lifetime in
Sec.~\ref{sec:broadening-mechanisms} and
\ref{sec:PhononLifetimeDiscussion}.
\label{sec:introPhononDOSprojection}

%%%%%%%%%%%%%%%%%%%%%%%%%%%%%%%%%%%%%%%%%%%%%%%%%%%%%%%%%%%%%%%%%%%%%%%%%%%%%%
\subsection{Calculation of the current}
\label{sec:CalculationOfCurrent} Our transport calculations are
based on NEGF techniques and in particular the Meir-Wingreen
formulation.\cite{MEWI.92.LANDAUERFORMULACURRENT,
JaWiMe.94.TIME-DEPENDENTTRANSPORTIN,Haug1996,Frederiksen2004} The
steady-state (spin-degenerate) electrical current $I_\alpha$ and the
power transfer $P_\alpha$ \emph{to} the device from lead
$\alpha=L,R$ can generally be expressed as
\begin{eqnarray}
\label{eq:MeirWingreenCurrent} I_\alpha &=& 2e\langle \dot {\widehat
N}_\alpha\rangle
    = \frac{-2e}\hbar \intR \frac{\intd \varepsilon}{2\pi} t_\alpha(\varepsilon),\\
P_\alpha &=& -2\langle \dot {\widehat H}_\alpha\rangle
    = \frac2\hbar \intR\frac{\intd \varepsilon}{2\pi}\varepsilon t_\alpha(\varepsilon), \label{eq:power}\\
t_\alpha(\varepsilon)&\equiv&\textrm{Tr}[\mathbf
    \Sigma^{<}_\alpha(\varepsilon)\mathbf G^{>}_D(\varepsilon)
    - \mathbf\Sigma^{>}_\alpha(\varepsilon) \mathbf G^{<}_D(\varepsilon)],\qquad \label{eq:tIntegrant}
\end{eqnarray}
where $\widehat N_\alpha$ is the electronic particle number operator
of lead $\alpha$, $\mathbf G^\lessgtr_D(\varepsilon)$ the full
lesser (greater) Green's function in the device region $D$
(including all relevant interactions), and $\mathbf
\Sigma^\lessgtr_\alpha(\varepsilon)$ the lesser (greater)
self-energy that represents the rate of electrons scattering into
(out of) the states in the device region $D$. We assume that the
leads are unaffected by the nonequilibrium conditions in the device
(this may be tested by increasing the device region). We can then
use the fluctuation--dissipation theorem to write the lead
self-energies as\cite{Haug1996}
\begin{eqnarray}
\mathbf
    \Sigma^{\lessgtr}_\alpha(\varepsilon) =\left\{\begin{array}{c}
    i n_\textrm{F}(\varepsilon-\mu_\alpha)\mathbf\Gamma_\alpha(\varepsilon)\\
    i [ n_\textrm{F}(\varepsilon-\mu_\alpha)-1] \mathbf\Gamma_\alpha(\varepsilon)\\
  \end{array}\right.,
\end{eqnarray}
where
$n_\textrm{F}(\varepsilon)=1/[\exp(\varepsilon/\text{k}_\text{B}T)+1]$
is the Fermi-Dirac distribution, $\mu_\alpha$ the chemical potential
of lead $\alpha$, and
\begin{eqnarray}
\mathbf\Gamma_\alpha(\varepsilon)\equiv
i[\mathbf\Sigma^r_\alpha(\varepsilon)
-\mathbf\Sigma^{a}_\alpha(\varepsilon)]=i[\mathbf\Sigma^>_\alpha(\varepsilon)
-\mathbf\Sigma^{<}_\alpha(\varepsilon)],
\end{eqnarray}
describes the broadening of the device states by the coupling to the
lead.

The lesser and greater Green's functions are generally related to
the retarded and advanced ones via the Keldysh equation
\begin{eqnarray}
\mathbf G^\lessgtr_D(\varepsilon)=\mathbf G^r_D(\varepsilon) \mathbf
\Sigma^\lessgtr_\textrm{tot}(\varepsilon) \mathbf
G^a_D(\varepsilon),
\end{eqnarray}
where $\mathbf \Sigma^\lessgtr_\textrm{tot}(\varepsilon)$ is the sum
of all self-energy contributions (leads, interactions, etc.).
Further, in steady-state situations time reversal symmetry relates
the advanced Green's function to the retarded one via $\mathbf
G^a_D(\varepsilon)=\mathbf
G^r_D(\varepsilon)^\dagger$.\cite{Haug1996}

%%%%%%%%%%%%%%%%%%%%%%%%%%%%%%%%%%%%%%%%%%%%%%%%%%%%%%%%%%%%%%%%%%%%%%%%%%%%%%
\subsection{Elastic transport}
\label{sec:ElasticTransport} If we consider a two-terminal setup
with no interactions in the device region $D$, then the current
expression simply reduces to the Landauer-B\"uttiker formula where
\Eqref{eq:tIntegrant} becomes
\begin{eqnarray}
t_\alpha(\varepsilon)&\equiv&
[n_\textrm{F}(\varepsilon-\mu_L)-n_\textrm{F}(\varepsilon-\mu_R)]\nonumber\\
&&\times\,\textrm{Tr}[
    \mathbf\Gamma_L(\varepsilon)\mathbf G^{0,r}_D(\varepsilon)
    \mathbf\Gamma_R(\varepsilon)\mathbf G^{0,a}_D(\varepsilon)].
\end{eqnarray}

{\transiesta} allows one to calculate the transmission function
$t(\varepsilon)=t_L(\varepsilon)=t_R(\varepsilon)$ under finite bias
conditions, i.e., with an electrostatic voltage drop over the device
and different chemical potentials of the two leads. Due to the
electrostatic self-consistency, this implies that the lead
self-energies, e.g.,~$\mathbf\Sigma^r_\alpha(\varepsilon)$, and
Hamiltonian $\mathbf H$ depend parametrically on the external bias
voltage $V$. These charging and polarization effects caused by the
electrostatic voltage
drop\cite{PaSt.01.Self-consistent-fieldstudyof} are fully treated in
{\transiesta} at finite bias. Although it is relatively
straightforward to include these effects, it is computationally
demanding for the inelastic calculation presented below. We have
therefore neglected the voltage dependence and used the zero-bias
self-energies and Hamiltonian in our inelastic calculations in the
low-bias regime. In the case of metallic leads and a small applied
bias (of the order of vibrational energies) we expect this
approximation to be accurate. However, sufficiently large biases
have been shown to influence the atomic
structure\cite{BrStTa.03.Originofcurrent-induced} as well as the
e-ph couplings.\cite{SeRoGu.05.Abinitioanalysis}

%%%%%%%%%%%%%%%%%%%%%%%%%%%%%%%%%%%%%%%%%%%%%%%%%%%%%%%%%%%%%%%%%%%%%%%%%%%%%%
\subsection{Self-consistent Born approximation}
\label{sec:SCBA} 
\begin{figure}[t!]
  \centering
  (a)
  \includegraphics[width=.3\columnwidth,angle=0]{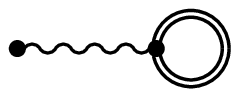}
  %\begin{fmffile}{DiagramHartree}
  %\begin{fmfgraph*}(70,40)
  %\begin{fmfsubgraph}(5,5)(60,30)
  %  \fmfleft{a0,l,a}
  %  \fmf{wiggly}{l,c}
  %  \fmf{double,left}{c,r,c}
  %  \fmfdot{l,c}
  %  \fmfright{b0,r,b}
  %\end{fmfsubgraph}
  %\end{fmfgraph*}
  %\end{fmffile}
  \quad
  (b)
  \includegraphics[width=.3\columnwidth,angle=0]{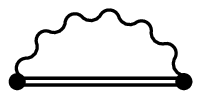}
  %\begin{fmffile}{DiagramFock}
  %\begin{fmfgraph*}(70,40)
  %\begin{fmfsubgraph}(5,5)(60,30)
  %  \fmfleft{lb,lt}
  %  \fmf{double}{rb,lb}
  %  \fmf{wiggly,left=.8,tension=.5}{lb,rb}
  %  \fmfdot{lb,rb}
  %  \fmfright{rb,rt}
  %\end{fmfsubgraph}
  %\end{fmfgraph*}
  %\end{fmffile}
  \caption{The lowest order diagrams for the phonon self-energies to the electronic
  description. The ``Hartree'' (a) and ``Fock'' (b) diagrams dress the electron
  Green's functions (double plain lines). The phonon Green's functions (single
  wiggly lines) are assumed to be described by the unperturbed ones, i.e., we
  ignore the e-ph renormalization of the phonon system.}
  \label{fig:Diagrams}
\end{figure}

Let us turn to the problem of the e-ph coupling. In
order to use \Eqref{eq:MeirWingreenCurrent} and (\ref{eq:power}) we
need the full Green's functions $\mathbf
G_D^{\lessgtr}(\varepsilon)$ taking the e-ph interaction into
account. Our approach is the SCBA where the phonon self-energy to
the electronic system is described by the diagrams shown in
\Figref{fig:Diagrams}.\cite{Haug1996} We note that in this work we
ignore the phonon renormalization (pair bubble diagram) by the e-ph
coupling.

We write the phonon self-energies from mode $\lambda$ as
\cite{Frederiksen2004,PaFrBr.05.Modelinginelasticphonona}
\begin{eqnarray}
\label{eq:SCBAselfenergies} \mathbf
\Sigma^{\lessgtr}_{\mathrm{ph},\lambda}(\varepsilon) &=&
    i \intR\frac{\intd\varepsilon'}{2\pi} \mathbf M^{\lambda}
    d_0^\lessgtr({\lambda},\varepsilon-\varepsilon')
    \mathbf G^\lessgtr_D(\varepsilon')\mathbf M^{\lambda},\nonumber\\\label{eq:SigmaLessGreat}\\
\mathbf \Sigma^{r}_{\mathrm{ph},\lambda}(\varepsilon) &=&
    \frac12 [\mathbf \Sigma^{>}_{\mathrm{ph},\lambda}(\varepsilon)
            -\mathbf
            \Sigma^{<}_{\mathrm{ph},\lambda}(\varepsilon)]\nonumber\\
            &&\quad-\frac i2 \mathcal H_{\varepsilon'}\{\mathbf
            \Sigma^{>}_{\mathrm{ph},\lambda}(\varepsilon')
            -\mathbf \Sigma^{<}_{\mathrm{ph},\lambda}(\varepsilon')\}(\varepsilon),\qquad\label{eq:SigmaRet}
\end{eqnarray}
where the retarded self-energy has been written in terms of the
lesser and greater self-energies using the Kramers-Kronig relation
$\mathcal H_{\varepsilon'}\{\mathbf G^r(\varepsilon')\}(\varepsilon)
= i \mathbf G^r(\varepsilon)$. The functional $\mathcal H$
represents the Hilbert transform described in \Appref{sec:Hilbert}.

The Hartree diagram \Figref{fig:Diagrams}(a) does not contribute to
the lesser and greater phonon self-energies; this is because energy
conservation implies that the wiggly line corresponds to a factor
$d^\lessgtr(\lambda,\varepsilon'=0)=0$.\cite{HYHEDA.94.RESONANT-TUNNELINGWITHELECTRON-PHONON}
It does, however, lead to constant term for the retarded self-energy
which can be understood as a static phonon-induced change in the
mean-field electronic potential.\cite{Haug1996,Frederiksen2004} From
\Eqref{eq:SigmaRet} we note that our retarded self-energy has the
limiting behavior $\lim_{\varepsilon\rightarrow \infty}\mathbf
\Sigma^{r}_{\mathrm{ph},\lambda}(\varepsilon)=0$. This is also the
limits of the Fock diagram \Figref{fig:Diagrams}(b) if one
calculates it directly with the Langreth
rules.\cite{Haug1996,Frederiksen2004} We therefore conclude that
\Eqref{eq:SigmaRet} gives exactly the Fock diagram. Ignoring the
Hartree term is reasonable since its small static potential shift
might be screened (at least partially) if it had been included on
the level of the DFT self-consistency loop. Further, the Hartree
diagram does not lead to a signal at the phonon threshold voltage.

The full device Green's functions $\mathbf
G_D^{r,\lessgtr}(\varepsilon)$ are related to $\mathbf
G_D^{0,r}(\varepsilon)$, $\mathbf
\Sigma_\alpha^{r,\lessgtr}(\varepsilon)$, and $\mathbf
\Sigma_{\mathrm{ph}}^{r,\lessgtr}(\varepsilon)\equiv
\sum_\lambda\mathbf
\Sigma_{\mathrm{ph},\lambda}^{r,\lessgtr}(\varepsilon)$ via the
Dyson and Keldysh equations \cite{Haug1996}
\begin{eqnarray}
\mathbf G_D^{r}(\varepsilon) &=& \mathbf G_D^{0,r}(\varepsilon)
+\mathbf G_D^{0,r} (\varepsilon) \mathbf
\Sigma^{r}_\mathrm{ph}(\varepsilon) \mathbf
G_D^{r}(\varepsilon),\label{eq:GRet}\\
\mathbf G_D^{\lessgtr}(\varepsilon) &=& \mathbf G_D^{r}(\varepsilon)
[\mathbf
\Sigma^{\lessgtr}_L(\varepsilon)+\mathbf\Sigma^{\lessgtr}_R(\varepsilon)
+\mathbf \Sigma^{\lessgtr}_\mathrm{ph}(\varepsilon)]\mathbf
G_D^{a}(\varepsilon).\label{eq:GLessGreat}\qquad
\end{eqnarray}
The coupled nonlinear
Eqs.~(\ref{eq:SigmaLessGreat})--(\ref{eq:GLessGreat}) have to be
solved iteratively subject to some constraint on the mode population
$\langle n_\lambda\rangle$ appearing in
$d^\lessgtr_0(\lambda,\varepsilon)$,
cf.~\Eqref{eq:92:PhGF0-lessergreater}. For weak e-ph coupling we
thus approximate the mode occupation $\langle n_\lambda\rangle$ by
the steady-state solution to a rate equation describing the heating
of the device
\begin{eqnarray}
\langle \dot n_\lambda\rangle &=& \frac{p_\lambda}{\hbar
\omega_\lambda}
    - \gamma^\lambda_\textrm{damp}[ \langle
    n_\lambda\rangle  - n_\textrm{B}(\hbar \omega_\lambda)],
\label{eq:RateEquation}
\end{eqnarray}
where
$n_\textrm{B}(\varepsilon)=1/[\exp(\varepsilon/\text{k}_\text{B}T)-1]$
is the Bose-Einstein distribution, $p_\lambda$ the power dissipated
into mode $\lambda$ by the electrons, and
$\gamma^\lambda_\textrm{damp}=1/\tau^\lambda_\textrm{ph}$ a damping
parameter related to the average lifetime of the phonon, e.g., by
coupling to bulk vibrations.

In steady state the power transferred by electrons from the leads
into to the device must balance the power transferred from the
device electrons to the phonons, i.e.,
\begin{eqnarray}
\label{eq:powerbalance} P_L+P_R = \sum_\lambda p_\lambda.
\end{eqnarray}
From the particle conservation condition\cite{Frederiksen2004}
\begin{eqnarray}
\label{eq:traceIdentity} \textrm{Tr}[\mathbf
    \Sigma^{<}_{\textrm{tot}}(\varepsilon)\mathbf G^{>}_D(\varepsilon)
    - \mathbf\Sigma^{>}_{\textrm{tot}}(\varepsilon) \mathbf
    G^{<}_D(\varepsilon)]=0,
\end{eqnarray}
we can define the quantity $p_\lambda$ as
\begin{eqnarray}
\label{eq:plambda-def} p_\lambda &\equiv& -\frac1\hbar
\intR\frac{\intd
\varepsilon}{2\pi}\,\varepsilon\\
&&\qquad\times\textrm{Tr}[\mathbf
    \Sigma^{<}_{\textrm{ph},\lambda}(\varepsilon)\mathbf G^{>}_D(\varepsilon)
    - \mathbf\Sigma^{>}_{\textrm{ph},\lambda}(\varepsilon) \mathbf
    G^{<}_D(\varepsilon)],\nonumber
\end{eqnarray}
which consequently obeys \Eqref{eq:powerbalance}. We note that in
this way we basically define $3N$ quantities from a single equation
for $\sum_\lambda p_\lambda$ only; different definitions could in
principle also fulfill the power balance. However, to lowest order
in the e-ph coupling our definition \Eqref{eq:plambda-def} is
unambiguously the power transferred to mode $\lambda$.

From \Eqref{eq:RateEquation} we can identify two regimes, (i) the
externally damped limit ($\gamma_\textrm{damp}^\lambda$ much larger
than electron-hole (e-h) pair damping $\gamma_\textrm{e-h}^\lambda$)
where the populations are fixed according to the Bose-Einstein
distribution $\langle n_\lambda\rangle
=n_\textrm{B}(\hbar\omega_\lambda)$, and (ii) the externally
undamped limit ($\gamma^\lambda_\textrm{damp}=0$ and hence from
\Eqref{eq:RateEquation} that $p_\lambda=0$) where the populations
vary with bias such that no power is dissipated in the device, i.e.,
$P_L+P_R=0$. It is instructive to note that $p_\lambda$ includes
both phonon emission and absorption processes, which is the reason
why a steady-state solution always exists.
\label{sec:DampingRegimes}

A typical situation that come close to the externally undamped limit
is when the device vibrations fall outside the phonon band of the
bulk electrodes, i.e., when there is a significant mass difference
between the device atoms and the electrode atoms. In this case the
vibrations cannot couple directly (resonantly) to the bulk, and the
damping, e.g., by anharmonic means, is likely to be much smaller
than the coupling to the electrons. One important example is the
hydrogen molecule clamped between platinum
contacts.\cite{SmNoUn.02.Measurementofconductance,DjThUn.05.Stretchingdependenceof}

To solve the SCBA equations
Eqs.~(\ref{eq:SigmaLessGreat})--(\ref{eq:RateEquation}), we have
developed an implementation in the programming language
\textsc{Python} where the Green's functions and self-energies are
sampled on a finite energy grid. The main technical challenges are
discussed in \Appref{sec:SCBAappendix}. Finally we note that with
the phonon self-energies
Eqs.~(\ref{eq:SigmaLessGreat})--(\ref{eq:SigmaRet}) the current is
conserved. This can be proven using the identity
\Eqref{eq:traceIdentity}.\cite{Frederiksen2004}

%%%%%%%%%%%%%%%%%%%%%%%%%%%%%%%%%%%%%%%%%%%%%%%%%%%%%%%%%%%%%%%%%%%%%%%%%%%%%%
\subsection{Lowest order expansion}

\label{sec:LOE} The solution of the SCBA equations is a daunting
numerical task for systems consisting of more than a handful of
atoms. However, for systems where the e-ph coupling is weak and the
density of states (DOS) varies slowly with energy, we have
previously derived the LOE
approximation.\cite{PaFrBr.05.Modelinginelasticphonona} Here we
elaborate on these results.

The main computational burden of the SCBA originates from the
numerical integration over energy needed in the evaluation of the
current and power expressions
Eqs.~(\ref{eq:MeirWingreenCurrent})--(\ref{eq:power}). The LOE
approximation assumes that the retarded and advanced single-particle
Green's functions $\mathbf G^{0,r/a}_D$ and lead self-energies
$\mathbf\Sigma^{r/a}_\alpha$ are energy \emph{independent}. We can
then expand the current and power expressions to the lowest order
(second) in e-ph couplings $\mathbf M^\lambda$ and perform the
energy integrations analytically. These integrals consist of
products of Fermi-Dirac functions and their Hilbert transforms. The
LOE thus retains the Pauli exclusion principle for fermionic
particles, which is necessary to model the blocking of phonon
emission processes at low bias.

In the LOE approximation, the total power dissipated into the phonon
system $P^\mathrm{LOE}\equiv P_L+P_R$ can, after lengthy
derivations, be written as \cite{PaFrBr.05.Modelinginelasticphonona}
\begin{eqnarray} \label{eq.power}
P^\mathrm{LOE}&=& \sum_\lambda p^\mathrm{LOE}_\lambda,\\
\label{eq:PowerLambda} p^\mathrm{LOE}_\lambda&=& \hbar
\omega_\lambda\left\{
    [n_\textrm{B}(\hbar\omega_\lambda)-\langle n_\lambda\rangle]\gamma_\textrm{e-h}^\lambda
    +\gamma_\textrm{em}^\lambda(V,T)\right\},\\
\label{eq:ElectronHolePairRate}
    \gamma_\textrm{e-h}^\lambda&=&\frac{\hbar\omega_\lambda}{\pi\hbar}
        \mathrm{Tr}\left[\mathbf{M}^\lambda\mathbf{A}\mathbf{M}^\lambda\mathbf{A}\right],\\
    \gamma_\textrm{em}^\lambda&=&\frac{
        \hbar\omega_\lambda[\cosh\!\big(\frac{e V}{\text{k}_\text{B}T}\!\big)\!-\!1]\coth\!\big(\frac{\hbar\omega_\lambda}{2\text{k}_\text{B}T}\!\big)\!
        -{eV}\sinh\!\big(\frac{e V}{\text{k}_\text{B}T}\!\big)\!}
        {\pi \hbar[\cosh\!\big( \frac{\hbar \omega_\lambda}{\text{k}_\text{B}T}\!\big)-\cosh\!\big( \frac{e V}{\text{k}_\text{B}T} \!\big)]}\nonumber\\
&&\times        \mathrm{Tr}\left[\mathbf{M}^\lambda
\mathbf{A}_L\mathbf{M}^\lambda\mathbf{A}_R\right],
\label{eq:EffectiveEmissionRate}
\end{eqnarray}
where the Bose-Einstein distribution $n_\textrm{B}(\varepsilon)$
appears in \Eqref{eq:PowerLambda} due to the integration of
Fermi-Dirac functions describing the electrons in the contacts. Here
$\mathbf{G}=\mathbf{G}^{0,r}_D(\varepsilon_\textrm{F})$,
$\mathbf{\Gamma}_{\alpha}=\mathbf{\Gamma}_{\alpha}(\varepsilon_\textrm{F})$,
and $\mathbf{A}=i(\mathbf{G}-\mathbf{G}^\dagger)$ are the
noninteracting retarded Green's function, the broadening by contact
$\alpha=L,R$, and the spectral function at $\varepsilon_\textrm{F}$,
respectively. For convenience we have also defined the quantities
$\mathbf{A}_\alpha=\mathbf{G}\mathbf{\Gamma}_{\alpha}\mathbf{G}^\dag$
such that $\mathbf{A}=\mathbf{A}_L+\mathbf{A}_R$.

The first term in \Eqref{eq:PowerLambda} describes the equilibrium
energy exchange between the vibrational and electronic degrees of
freedom (e-h pair damping $\gamma_\textrm{e-h}^\lambda$ of the
vibrations); it tend to drive the phonon system towards the
Bose-Einstein distribution. The second term appears in
nonequilibrium and is related to an effective emission rate
$\gamma_\textrm{em}^\lambda$ of vibrational quanta under finite
bias. At low temperatures ($\text{k}_\text{B}T\rightarrow 0$) this
rate is given as
\begin{eqnarray}
\gamma_\textrm{em}^\lambda&=&
  \frac{|eV|-\hbar\omega_\lambda}{\pi\hbar} \,\theta(|eV|-\hbar\omega_\lambda)\mathrm{Tr}\left[\mathbf{M}^\lambda\mathbf{A}_L\mathbf{M}^\lambda\mathbf{A}_R\right],\nonumber\\
\end{eqnarray}
where $\theta(x)$ is the step function; i.e., the net emission of
phonons above the threshold grows linearly with the bias voltage.
Furthermore, since
$\mathrm{Tr}\left[\mathbf{M}^\lambda\mathbf{A}_\alpha
\mathbf{M}^\lambda\mathbf{A}_\beta \right]\geq 0$, we find that
\begin{eqnarray} \label{eq:TraceInequality} \mathrm{Tr}\left[
\mathbf{M}^\lambda \mathbf{A} \mathbf{M}^\lambda \mathbf{A} \right]
\geq 2 \,\mathrm{Tr}\left[\mathbf{M}^\lambda\mathbf{A}_L
\mathbf{M}^\lambda\mathbf{A}_R \right].
\end{eqnarray}
We can use this inequality to derive an upper bound on the phonon
occupation by solving the steady-state condition
$p^\mathrm{LOE}_\lambda=0$ [cf.~\Eqref{eq:RateEquation} with no
external damping]. It simply
becomes\cite{MiTiUe.02.Theoryofvibrational,MiTiUe.03.Spectralfeaturesof}
\begin{eqnarray}
\label{eq:occupationBound} \langle n_\lambda\rangle &\leq& \frac 12
\frac{|eV|-\hbar\omega_\lambda}{\hbar\omega_\lambda}\,
\theta(|eV|-\hbar\omega_\lambda).
\end{eqnarray}

\begin{figure}[t!]
\begin{center}
%\psfrag{m1}[c][c]{$\mu_L$} \psfrag{m2}[c][c]{$\mu_R$}
%\psfrag{A}[c][c]{a)} \psfrag{B}[c][c]{b)} \psfrag{C}[c][c]{c)}
%\psfrag{D}[c][c]{d)} \psfrag{E}[c][c]{E} \psfrag{V}[c][c]{$eV$}
%\psfrag{Vmhw}[c][c]{$e V-\hbar \omega$} \psfrag{Vphw}[c][c]{$e5V+\hbar \omega$} \psfrag{hw}[c][c]{$\hbar \omega$}
%\psfrag{O1}[c][c]{Left} \psfrag{O2}[c][c]{Right}
%\includegraphics[width=\columnwidth,angle=0]{./figures/LOEfig.eps}
%\includegraphics[width=\columnwidth,angle=0]{./figures/LOEfig_psfragconverted.eps}
\includegraphics[trim = 80 0 0 0,width=\columnwidth]{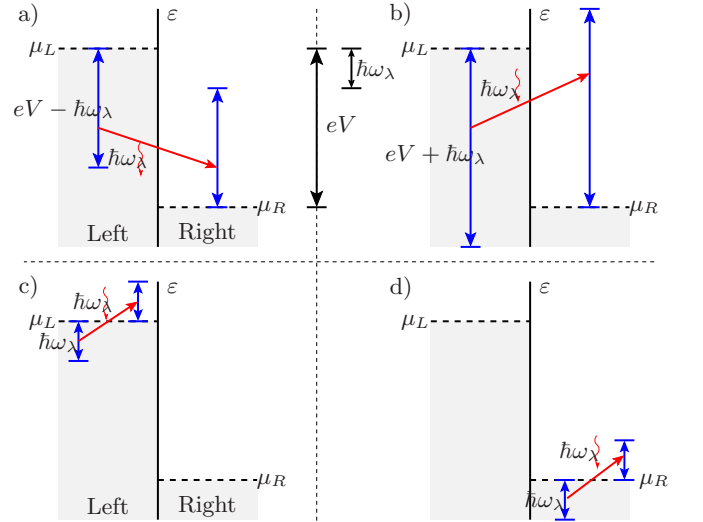}
\end{center}
\caption{Schematic representation of the energy phase space
available for scattering processes due to the Pauli principle.
Phonon emission (a) and absorption (b) between scattering states
originating from the left and right contacts. Figs.~(c) and (d)
correspond to phonon absorption between scattering states in the
same contact. \label{fig:LOEphaseFig} }
\end{figure}

To provide an intuitive understanding of
Eqs.~(\ref{eq.power})--(\ref{eq:occupationBound}) consider the
following arguments: The energy phase space available for phonon
emission and absorption processes is limited by the Pauli principle,
as sketched in \Figref{fig:LOEphaseFig}. We divide the electronic
phase space in two, corresponding to scattering states incoming from
either the left or the right contact. Without e-ph scattering these
states are assumed to be populated up to the Fermi level
$\varepsilon_\textrm{F}$ (we take $\mu_L>\mu_R+\hbar\omega_\lambda$
and $\text{k}_\text{B}T\rightarrow 0$). Within this picture phonon
emission can only take place from a populated state originating in
the left contact to an empty state originating in the right contact,
see \Figref{fig:LOEphaseFig}(a). Similarly, phonon absorption can be
described by three different processes sketched in
\Figref{fig:LOEphaseFig}(b)--(d), again corresponding to scattering
from populated initial states to empty final states.

The scattering rates for these processes are proportional to the
energy window in which they can take place. Denoting the scattering
rate per energy as $d\gamma_{\alpha\alpha'}/d\varepsilon$, where
$\alpha=L,R$ ($\alpha'=L,R$) indicates the propagation direction of
the initial (final) scattering state, we can write the spontaneous
plus stimulated emission power as
$p^\mathrm{LOE}_{\lambda,\textrm{em}} = \hbar \omega_\lambda
(\langle n_\lambda\rangle+1) \left(eV-\hbar \omega_\lambda\right)
d\gamma_{LR}/d\varepsilon$ and the absorption power as
$p^\mathrm{LOE}_{\lambda,\textrm{ab}} = \hbar \omega_\lambda \langle
n_\lambda\rangle [ \left(eV+\hbar \omega_\lambda\right)
d\gamma_{LR}/d\varepsilon$ $+ \hbar\omega_\lambda \left(
d\gamma_{LL}/d\varepsilon+d\gamma_{RR}/d\varepsilon\right) ]$. The
net power transfer from the electronic system to the phonon mode
$\lambda$ is therefore
\begin{eqnarray}
\label{eq:simplePowerArgument}
p^\mathrm{LOE}_{\lambda} &=& p^\mathrm{LOE}_{\lambda,\textrm{em}}-p^\mathrm{LOE}_{\lambda,\textrm{ab}}\nonumber \\
&=&-(\hbar \omega_\lambda)^2 \langle n_\lambda\rangle \left[ 2
\frac{d\gamma_{LR}}{d\varepsilon} +
\frac{d\gamma_{LL}}{d\varepsilon}
+\frac{d\gamma_{RR}}{d\varepsilon} \right]\nonumber\\
&&  + \hbar \omega_\lambda  \, (eV-\hbar \omega_\lambda) \,
\frac{d\gamma_{LR}}{d\varepsilon}.
\end{eqnarray}
A comparison with \Eqref{eq:PowerLambda} reveals that the term
proportional to the occupation $\langle n_\lambda\rangle$ is bias
independent due to a cancelation of phonon absorption by stimulated
emission. Furthermore, the upper bound in \Eqref{eq:occupationBound}
is directly motivated by equating \Eqref{eq:simplePowerArgument} to
zero (steady state) and by ignoring scattering processes with
initial and final states propagating in the same direction
($d\gamma_{\alpha\alpha}/d\varepsilon$). In addition, a steady-state
solution to \Eqref{eq:RateEquation} always exists because the phonon
emission rate is always smaller than the total phonon absorption
rate, and that emission processes are restricted to a smaller energy
window than absorption processes.

The LOE approximation, which above was applied to the power, also
allows us to write the current through the device $I^\mathrm{LOE}$
as\cite{PaFrBr.05.Modelinginelasticphonona,ViCuPa.05.Electron-vibrationinteractionin}
\begin{widetext}
\begin{eqnarray}
 I^\mathrm{LOE} &=&   \textrm{G}_0 V \mathrm{Tr}
\left[\mathbf{G} \mathbf{\Gamma}_R \mathbf{G}^\dag
\mathbf{\Gamma}_L\right]  \nonumber \\ & & + \mysum{\lambda}{} {\cal
I}^\mathrm{sym}_\lambda(V,T, \langle n_\lambda\rangle) \,
   \mathrm{Tr}\left[
     \mathbf{G}^\dag \mathbf{\Gamma}_L \mathbf{G} \left\{
     \mathbf{M}^\lambda \mathbf{A}_R  \mathbf{M}^\lambda +
       \frac{i}{2}\left( \mathbf{\Gamma}_R \mathbf{G}^\dag \mathbf{M}^\lambda \mathbf{A} \mathbf{M}^\lambda - \mathrm{h.c.}\right)
       \right\}
   \right]  \nonumber \\
 & & +
\mysum{\lambda}{} {\cal I}^\mathrm{asym}_\lambda(V, T)\,
  \mathrm{Tr}\left[
         \mathbf{G}^\dag \mathbf{\Gamma}_L \mathbf{G} \left\{
           \mathbf{\Gamma}_R \mathbf{G}^\dag \mathbf{M}^\lambda
           \left( \mathbf{A}_R-\mathbf{A}_L \right) \mathbf{M}^\lambda +
           \mathrm{h.c.}
           \right\}
    \right] ,
 \label{eq:current1} \\
{\cal I }^\mathrm{sym}_\lambda & = &
      \frac{e }{\pi \hbar} \mybpar{{2 e V} \langle n_\lambda \rangle +
      \frac{\hbar \omega_\lambda-{e V}}{e^{\frac{\hbar \omega_\lambda-e V}{\text{k}_\text{B}T}}-1}-
      \frac{{\hbar \omega_\lambda}+{e V}}{e^{\frac{\hbar \omega_\lambda+e V}{\text{k}_\text{B}T}}-1}} ,
\label{eq:currentNormal} \\
 {\cal I}^{\mathrm{asym}}_\lambda&=&
    \frac{e}{\hbar}\myint{-\infty}{\infty} \frac{d\varepsilon}{2\pi} \left[n_\textrm{F}(\varepsilon) -n_\textrm{F}(\varepsilon-e V)\right] \,
    {\cal H}_{\varepsilon'} \{{n_\textrm{F}(\varepsilon'+\hbar \omega_\lambda)
    -n_\textrm{F}(\varepsilon'-\hbar \omega_\lambda)}\}(\varepsilon) ,
   \label{eq:currentHilbert}
\end{eqnarray}
\end{widetext}
where the bias is defined via $eV=\mu_R-\mu_L$, and
$\textrm{G}_0=2e^2/h$ is the spin-degenerate conductance quantum.
This expression is current conserving, i.e., calculating the current
at the left and right contacts give the same result.

\begin{figure}[t!]
  \centering
  \includegraphics[width=\columnwidth]{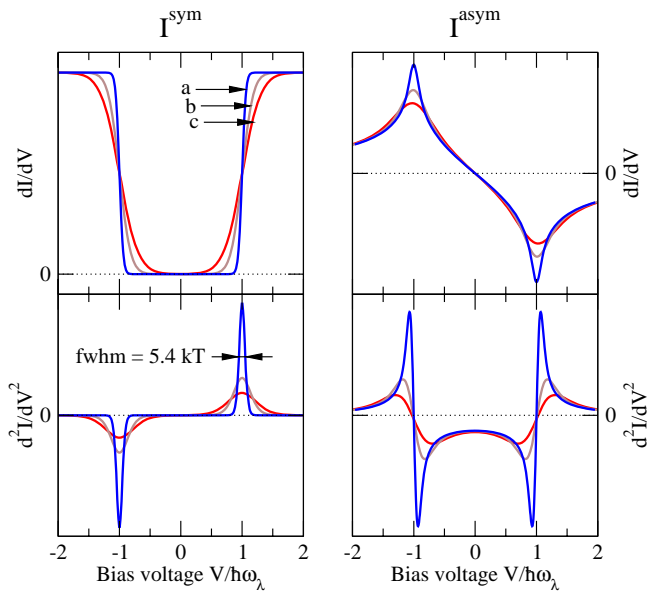}
  \caption{(Color online) Universal functions
  \Eqref{eq:currentNormal} and (\ref{eq:currentHilbert}) giving
  symmetric and asymmetric phonon contributions to the conductance in the LOE, respectively. The
  differential conductance $dI/dV$ and the second derivative $d^2I/dV^2$ are
  shown (in arbitrary units) for one phonon mode for three different
  temperatures (a) $\text{k}_\text{B}T/\hbar\omega_\lambda=0.02$, (b) $\text{k}_\text{B}T/\hbar\omega_\lambda=0.06$,
  and (c) $\text{k}_\text{B}T/\hbar\omega_\lambda=0.10$.}
  \label{fig:LOEshape}
\end{figure}

The LOE expression for the current \Eqref{eq:current1} contains
three terms, (i) the Landauer-B\"uttiker term corresponding to the
elastic conductance, (ii) the ``symmetric'' term corresponding to
symmetric conductance steps at the vibrational energies, and (iii)
the ``asymmetric'' term corresponding to peaks and dips in the
conductance which are asymmetric with voltage inversion, see
\Figref{fig:LOEshape}. For geometrically symmetric junctions, it can
be shown that the asymmetric term vanishes exactly. Even for
geometrically asymmetric systems we typically find that it is a very
small contribution compared with the symmetric term. Furthermore,
the sign of the conductance step for the symmetric term in general
shows an increase (decrease) in the conductance for low (high)
conducting systems, e.g., vibrations usually help electrons through
molecules while they backscatter electrons in atomic wires. This is
discussed further for a one-level model in
Ref.~\onlinecite{PaFrBr.06.Phononscatteringin}.

The LOE approximation is computationally simple and can be applied
to systems of considerable size. Although the approximation is not
strictly valid for systems with energy-dependent DOS, comparison
with the full SCBA calculations shows good agreement even for
systems that have a slowly varying DOS (on the scale of vibrational
energies), e.g., the organic molecules connected to gold electrodes
described below in \Secref{sec:HydrocarbonMolecules}. The LOE
approximation will certainly fail when sharp resonances (compared to
the vibrational energies) are present within the order of phonon
energies of the Fermi energy. However, in this case Coulomb blockade
physics is also expected, which thus makes any DFT mean-field
approach (including ours) questionable.

%%%%%%%%%%%%%%%%%%%%%%%%%%%%%%%%%%%%%%%%%%%%%%%%%%%%%%%%%%%%%%%%%%%%%%%%%%%%%%
\subsection{Broadening mechanisms}
\label{sec:broadening-mechanisms} The width of the experimentally
measured phonon signal in the conductance is a combination of (at
least) three broadening mechanisms, namely the intrinsic ones from a
finite temperature and a finite phonon lifetime, as well as the one
related to the modulation voltage used in lock-in measurements (to
improve the signal-to-noise ratio) of the second derivative of the
current with respect to the bias. These contributions do not add up
trivially. However, as we show below, one can provide estimates for
each of the different contributions which thus help to understand
what effect is the dominant one.

As can be seen in \Figref{fig:LOEshape}, the electronic temperature
gives rise to a broadening of the vibrational signal. From
\Eqref{eq:currentNormal} the full width half max (FWHM) in the
second derivative of the current can be shown to be approximately
$5.4~\text{k}_\text{B}T$.\cite{LAJA.68.MOLECULARVIBRATIONSPECTRA,
HA.77.INELASTICELECTRON-TUNNELING,PaFrBr.05.Modelinginelasticphonona}

The effects of a finite phonon lifetime
$\tau_\textrm{ph}^\lambda=1/\gamma^\lambda_\textrm{damp}$ is to a
first approximation described by a convolution of the free phonon
Green's functions with a Lorentzian with a HWHM width of
$\hbar\gamma^\lambda_\textrm{damp}$. Consequently, this convolution
propagates to the phonon self-energies \Eqref{eq:SCBAselfenergies}
and to the inelastic LOE corrections to the current,
cf.~\Eqref{eq:currentNormal} and (\ref{eq:currentHilbert}). The FWHM
broadening in the second derivative of the current is thus
$2\hbar\gamma_\textrm{damp}$. The intrinsic linewidth of the phonon
signal has also been discussed in a simple SCBA model by Galperin
\etal\cite{GaRaNi.04.Onlinewidths}

The broadening from the lock-in technique for measurements of the
first or second derivatives of the current can be estimated in the
following way. With a small harmonic modulation signal (with
amplitude $A=\sqrt{2} \,V_{\mathrm{rms}}$) applied on top of the
bias voltage one can measure derivatives of the current. As shown in
\Appref{sec:ddIlockin} the FWHM width induced by the lock-in
measurement technique is $2.45 \, V_{\mathrm{rms}}$ and $1.72 \,
V_{\mathrm{rms}}$ for the first and second derivatives of the
current, respectively (neglecting intrinsic broadening). In other
words, if $d^2I/dV^2$ is a $\delta$-function, the experimentally
measured FWHM width will be either $2.45 \, V_{\mathrm{rms}}$ or
$1.72 \, V_{\mathrm{rms}}$, depending on whether the lock-in
measurement is on the first or second harmonic.

%%%%%%%%%%%%%%%%%%%%%%%%%%%%%%%%%%%%%%%%%%%%%%%%%%%%%%%%%%%%%%%%%%%%%%%%%%%%%%%%%%%%%
%\input{AtomicGoldWires}
%%%%%%%%%%%%%%%%%%%%%%%%%%%%%%%%%%%%%%%%%%%%%%%%%%%%%%%%%%%%%%%%%%%%%%%%%%%%%%%%%%%%%
%%%%%%%%%%%%%%%%%%%%%%%%%%%%%%%%%%%%%%%%%%%%%%%%%%%%%%%%%%%%%%%%%%%%%%%%%%%%%%
\section{Atomic gold wires}
\label{sec:AtomicGoldWires}
\begin{figure}[t!]
    \centering
  %\includegraphics[width=\columnwidth]{figures/NPQRSvertical_L.eps}
  %\begin{picture}(0,0)
  %  \put(-110,2){(a)}
  %  \put(-58,2){(b)}
  %  \put(-8,2){(c)}
  %  \put(44,2){(d)}
  %  \put(85,2){(e)}
  %  \put(100,60){$L$}
  %\end{picture}
  \includegraphics[width=\columnwidth]{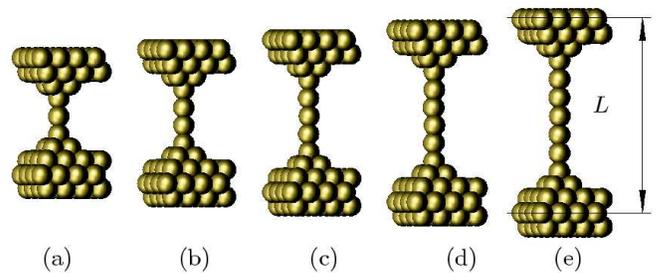}
  \caption{(Color online) Generic gold wire supercells containing 3 to 7 atoms bridging
  pyramidal bases connected to stacked Au(100) layers.
  As indicated on the figure, the electrode separation $L$ is defined
  as the distance between the plane in each electrode containing the
  second-outermost Au(100) layer.}
  \label{fig:Structures}
\end{figure}

Since the discovery in the late 1990s that gold can form
free-standing wires of single atoms
\cite{FiLyMc.97.Atomisticsimulationof,SoBrJa.98.Mechanicaldeformationof,
OhKoTa.98.Quantizedconductancethrough,YaBova.98.Formationandmanipulation}
the mechanical, chemical, and electrical properties of these
atomic-scale systems have been extensively
studied.\cite{ScAgCu.98.signatureofchemical,
SaArJu.99.Stiffmonatomicgold,ToToDa.99.puzzlingstabilityof,
EmKi.99.Electronstanding-waveformation,
ToHoSu.00.Current-inducedforcesin,HaBaSc.00.Nanowiregoldchains:,
SmUnYa.01.Commonoriginsurface,BaJa.01.Chainformationof,
RuBaAg.01.Mechanicalpropertiesand,
SaArJu.01.Zigzagequilibriumstructure,dadaFa.01.Howdogold,
NiBrHa.02.Current-voltagecurvesof,UnYaGr.02.Calibrationoflength,
AgUnRu.02.Onsetofenergy,AgUnRu.02.Electrontransportand,
MoHoTo.03.Inelasticcurrent-voltagespectroscopy,
SmUnRu.03.Observationofparity,SmUnva.04.high-biasstabilityof,
daNoda.04.Theoreticalstudyof,ZhEr.04.Zero-voltageconductanceof,
FrBrLo.04.InelasticScatteringand,DrPaHe.05.Structureandconductance}
For this reason we illustrate in this section our method described
in \Secref{sec:ElectronicStructureMethods} and \ref{sec:Transport}
by applying it to model inelastic scattering in atomic gold wires.
We compare directly the results of our theoretical developments with
the high-quality experimental data by Agra\"it and
co-workers.\cite{AgUnRu.02.Onsetofenergy,AgUnRu.02.Electrontransportand}
They used a cryogenic STM to first create an atomic gold wire
between the tip and the substrate surface, and then to measure the
conductance against the displacement of the tip. From the length of
the observed conductance plateau around G$_0$---the signature that
an atomic wire has been formed---it was possible to determine the
approximate size as well as the level of strain of the created wire.
Under these conditions Agra\"it \etal~then used point-contact
spectroscopy to show that the conductance of an atomic gold wire
decreases a few percent around a particular tip-substrate voltage
(symmetric around zero bias) presumably coinciding with the natural
frequency of a certain vibrational mode of the wire. With this
inelastic spectroscopy method they could further characterize the
conductance drop as a function of wire length and strain.

To simulate these experiments, we study wires containing different
number of atoms and under varying stretching conditions. The generic
supercells used in the {\siesta} calculations are illustrated in
\Figref{fig:Structures} and consist of 3 to 7 gold atoms bridging
pyramidal bases connected to stacked Au(100) layers. We use a
$4\times 4$ supercell size in the plane transverse to the transport
direction and define the electrode separation $L$, as indicated on
\Figref{fig:Structures}, as the distance between the plane in each
electrode containing the second-outermost Au(100) layer. The
face-centered cubic (FCC) lattice constant for the bulk gold atoms
is taken to be $a=4.18$ {\AA}.\footnote{The FCC lattice constant for
Au is theoretically $a=4.18$ {\AA} in a high quality plane-wave DFT
calculation and $a=4.21$ {\AA} in a {\siesta} calculation with our
present DFT settings. The experimental value is $a=4.05$ {\AA}.}

We generally use (unless otherwise specified) the
Perdew-Burke-Ernzerhof version of the GGA for the
exchange-correlation
functional,\cite{PeBuEr.96.Generalizedgradientapproximation} a
split-valence single-$\zeta$ plus polarization (SZP) basis set with
a confining energy of 0.01 Ry (nine orbitals corresponding to the
$5d$ and $6(s,p)$ states of the free Au atom), a cutoff energy of
200 Ry for the real space grid integration, and the $\Gamma$-point
approximation for the sampling of the three-dimensional Brillouin
zone. The interaction between the valence electrons and the ionic
cores are described by a standard norm-conserving Troullier-Martins
pseudopotential\cite{TrMa.91.EfficientPseudopot} generated from a
relativistic atomic calculation (including core correction). We have
found that these settings yield a reasonable compromise between
accuracy and computational cost.

\begin{figure}[t!]
  \centering
  \includegraphics[width=.87\columnwidth]{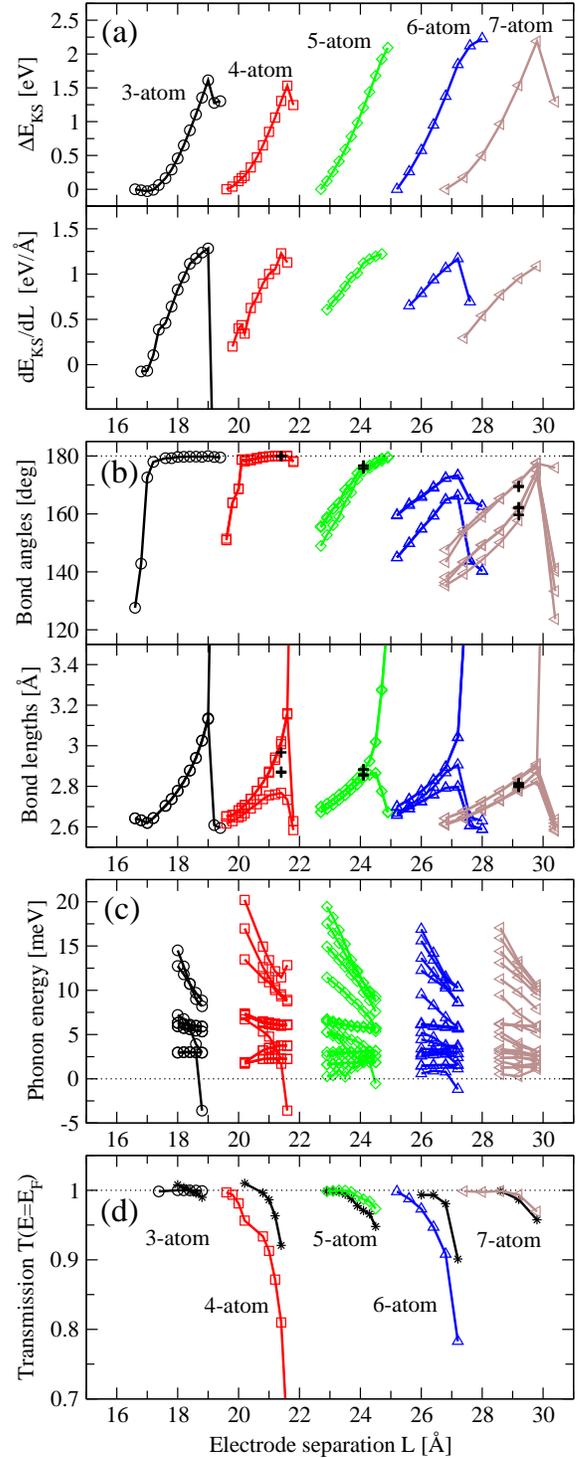}
  \caption{(Color online) Energetic, geometric, and conductive properties of atomic gold
  wires: (a) Kohn-Sham total energy (cohesive energy) vs.~electrode
  separation, (b) bond angles and bond lengths, (c) phonon energies, and (d)
  elastic transmission at the Fermi energy calculated both
  for the $\Gamma$-point (colored open symbols) as
  well as with a $5\times 5$ $\mathbf{k}$-point sampling of
  the two-dimensional Brillouin zone perpendicular to the transport
  direction (black stars).}
  \label{fig:EtotVsLength}
\end{figure}

\subsection{Geometry relaxation}
For a given electrode separation $L$ the first calculational step is
to relax the geometry to obtain a local energy minimum configuration
$\mathbf R^0$. With the settings described above we relax both the
outermost electrode layers, the pyramidal bases, and the wire atoms
until all forces acting each of these atoms are smaller than
$F_\textrm{max}=0.02$ {eV/\AA}.

Figure~\ref{fig:EtotVsLength}(a) shows the relative differences in
the Kohn-Sham total energy (cohesive energy) as the wires are
elongated. We also show the numerical derivatives of these binding
energy curves as a measure of the forces acting on the wire. The
breaking force, defined as the energy slope of the last segment
before breaking, is found be of the order 1 eV/{\AA} $\sim 1.6$ nN.
This agrees well with the experimental results which have shown the
break force for atomic gold wires to be close to 1.5
nN.\cite{BaJa.01.Chainformationof,RuBaAg.01.Mechanicalpropertiesand,
AgYeva.03.Quantumpropertiesof}

In \Figref{fig:EtotVsLength}(b) we summarize the geometrical
findings of the relaxation procedure by plotting the wire bond
lengths and bond angles as a function of electrode separation $L$.
The figure shows that the short wires containing 3 or 4 atoms adopt
a linear structure over a wide range of electrode separations. The
longer wires, on the other hand, are generally found to have a
zigzag geometry only approaching a linear form when they are
stretched close to the breaking
point.\cite{SaArJu.99.Stiffmonatomicgold}

From the plot of the bond lengths between nearest neighbors in the
wire one notices that the 4 and 6 atom wires have a more pronounced
tendency to dimerize than the wires with an odd number (due to
left/right symmetry of the structures only wires with an even number
of atoms should be able to dimerize). In three test calculations
with a $3\times 3\times 3$ $\mathbf{k}$-point sampling of the three
dimensional Brillouin zone we generally achieve very similar atomic
arrangements as compared to the $\Gamma$-point only. However, these
calculations, which are indicated with black crosses in
\Figref{fig:EtotVsLength}(b), seem to reduce the dimerization
tendency somewhat.

\subsection{Vibrational analysis}
\label{subsec:VibrationalAnalysis} We calculate the vibrational
frequencies and modes as described in \Secref{sec:VibModes}. With
$N$ vibrating atoms we thus find $3N$ modes for a given structure.
The phonon spectrum for the wire is plotted in
\Figref{fig:EtotVsLength}(c), where negative values indicate modes
with imaginary frequency implying the breaking of an unstable wire.
The general trend is that the phonon energies diminish as the wires
are elongated. This can be understood by considering that the
effective ``springs'' between ions in the wires are softened as the
bonds are stretched, which in turn result in lower energies.

In the results to follow we generally take the wire and pyramidal
base atoms as the dynamic region (as indicated in
\Figref{fig:GenericTransportSetup}), i.e.~these atoms are allowed to
vibrate. For the 3- to 7-atom wires this leave us with 33 to 45
vibrational modes. The corresponding e-ph couplings are calculated
in a slightly larger device region containing also the outermost
surface layer. This inclusion of an extra layer is necessary to
represent the vibrational modulation of the hopping between the
pyramidal base atoms and the first surface layers.

\subsection{Elastic transmission}
\label{sec:elastictransmission}

\begin{figure}[t!]
  \centering
  \includegraphics[width=\columnwidth]{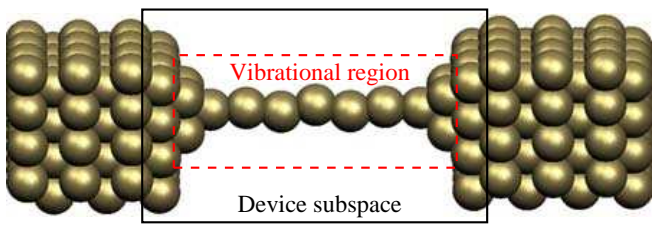}
  \caption{(Color online) Generic transport setup in which a relaxed
  wire geometry---here a 7-atom wire with $L=29.20$ {\AA}---is coupled to
  semi-infinite electrodes. As indicated on the
  figure the vibrational region is taken to include the atoms in the pyramidal
  bases and the wire itself, whereas the device region (describing the e-ph couplings)
  includes also the outermost surface layers.}
  \label{fig:GenericTransportSetup}
\end{figure}

In order to determine the transport properties of the wire
geometries described above, we construct from the supercells shown
in \Figref{fig:Structures} new wire geometries which are coupled to
semi-infinite electrodes as schematically illustrated in
\Figref{fig:TwoGenericUnitCells}(b). The resulting setup is shown in
\Figref{fig:GenericTransportSetup} for the case of a 7-atom long
gold wire. As indicated on this figure we consider the device
subspace to include the top-most surface layer, the pyramidal bases,
and the wire itself.

The elastic transmission evaluated at the Fermi energy
$\varepsilon_\textrm{F}$ is calculated using {\transiesta} described
in
Ref.~\onlinecite{BrMoOr.02.Density-functionalmethodnonequilibrium}.
The results are shown in \Figref{fig:EtotVsLength}(d) both for the
$\Gamma$-point (open symbols) as well as with a $5\times 5$
$\mathbf{k}$-point sampling of the two-dimensional Brillouin zone
perpendicular to the transport direction (black stars). In
correspondence with previous work,
e.g.,~Refs.~\onlinecite{ScAgCu.98.signatureofchemical,BrKoTs.99.Conductionchannelsat,
EmKi.99.Electronstanding-waveformation,
SmUnRu.03.Observationofparity}, we find that the total transmission
is close to unity, except for the very stretched configurations
where the transmission goes down somewhat. From
\Figref{fig:EtotVsLength}(d) one observes a reasonable agreement
between the $\Gamma$-point and the $\mathbf{k}$-point sampled
transmissions, particularly when the transmission is close to one.
Worst are the discrepancies for the 4 and 6 atom wires, which also
are the cases where the transmission deviates most from unity. We
subscribe these signatures to the so-called odd-even behavior in the
conductance of metallic atomic wires, in which perfect transmission
is expected only for an odd number of atoms in a chain. For an even
number of atoms the conductance should be
lower.\cite{AgYeva.03.Quantumpropertiesof} Further, the observed
dimerization is also expected to reduce the conductance (the Peierls
instability for infinite metallic wires results in the opening of a
band gap at the Fermi energy). We also note that on an energy scale
of the typical phonon energies the transmission function is to a
very good approximation a constant around the Fermi energy.

\subsection{Inelastic transport}
\label{sec:INelastictransmission}

Having determined the vibrational frequencies, the e-ph couplings,
and the elastic transmission properties, we are in position to
calculate the inelastic current as described in
\Secref{sec:CalculationOfCurrent}.

\begin{figure}[t!]
  \centering
  \includegraphics[width=0.85\columnwidth]{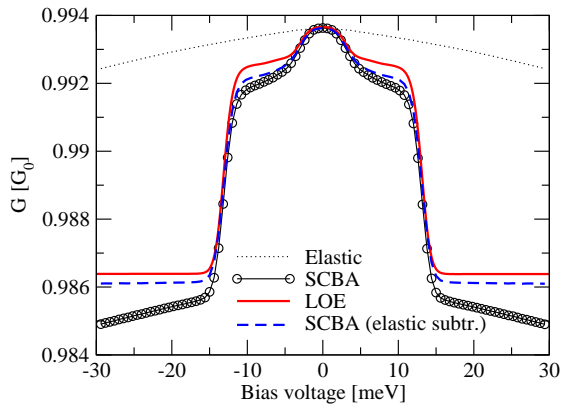}
  \caption{(Color online) Elastic and inelastic differential conductance calculated
  at $T=10.0$ K in a reduced device region for the 7-atom wire shown in
  \Figref{fig:GenericTransportSetup}. The small variation in elastic conductance
  with bias (dotted curve) relates to a weak energy dependence of the elastic
  transmission function at the $\Gamma$-point around $\varepsilon_\textrm{F}$.
  The full SCBA calculation (circles)  follows this trend and shows on top of it
  symmetric drops characteristic for phonon scattering. The LOE
  calculation (line) does not include the elastic variation but
  gives basically the same predictions for the inelastic signals as the SCBA
  with the elastic background signal subtracted (dashed curve).
  This illustrates the agreement between the LOE and SCBA approaches
  for the inelastic contribution.}
  \label{fig:SCBAvsLOEforAuWires}
\end{figure}

\begin{figure*}[t!]
  \centering
  \includegraphics[width=\textwidth]{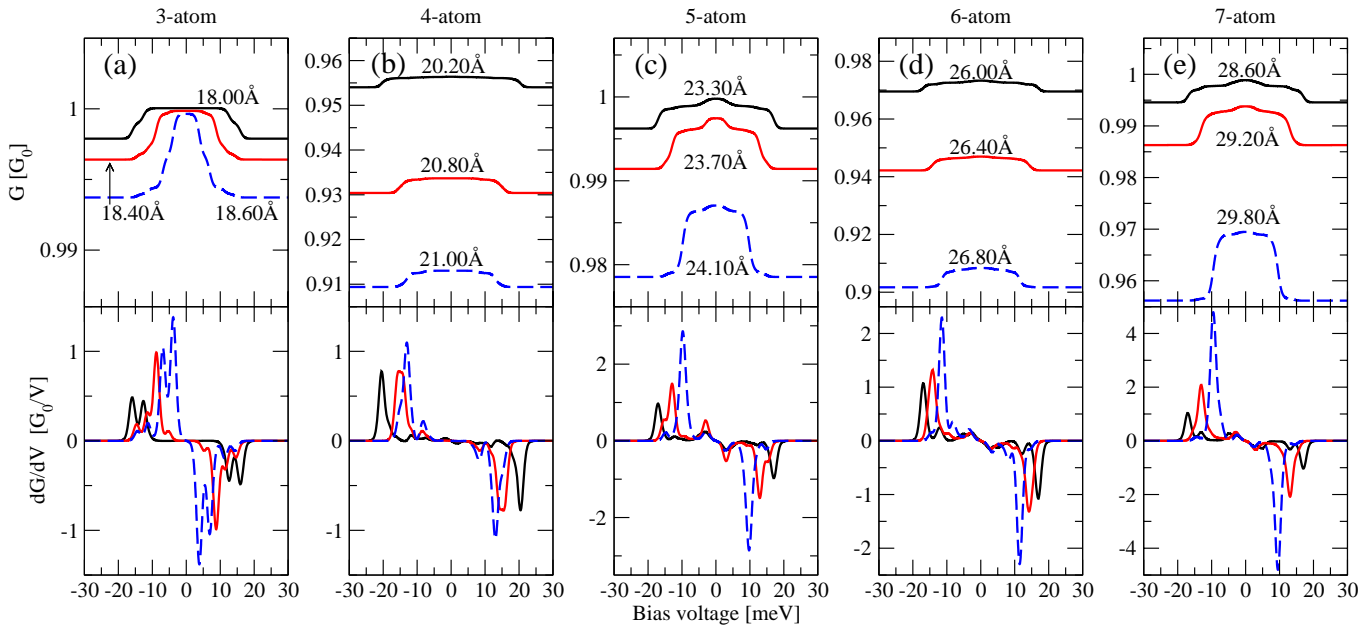}
  \caption{(Color online) The differential conductance $G$ and its derivative $dG/dV$ calculated with
  the LOE approach for the 3- to 7-atom gold wires in the externally damped limit.
  The electrode separation $L$ is indicated next to the conductance curves. As shown in
  \Figref{fig:GenericTransportSetup} the device region includes the outermost electrode
  layer whereas the dynamic atoms are pyramidal bases plus wire. The temperature of the
  leads is $T=4.2$ K.}
  \label{fig:LOEcalculationsForAu}
\end{figure*}

\begin{figure*}[t]
  \centering
  %\includegraphics[width=\columnwidth]{figures/SignalsAndStrengthVsL_v2.eps}\hspace{3mm}
  %\includegraphics[width=\columnwidth]{figures/ZcomponentVsL_v3.eps}\\
  %\vspace{6mm}
  %\includegraphics[width=\columnwidth]{figures/ABLcomponentVsL_v3.eps}\hspace{3mm}
  %\includegraphics[width=\columnwidth]{figures/LocalizationInChainVsL_v3.eps}\\\vspace{-3mm}
  %\begin{picture}(0,0)(0,0)
  %\put(-245,320){(a)}
  %\put(10,320){(b)}
  %\put(-245,150){(c)}
  %\put(10,150){(d)}
  %\end{picture}
  \includegraphics[width=\textwidth]{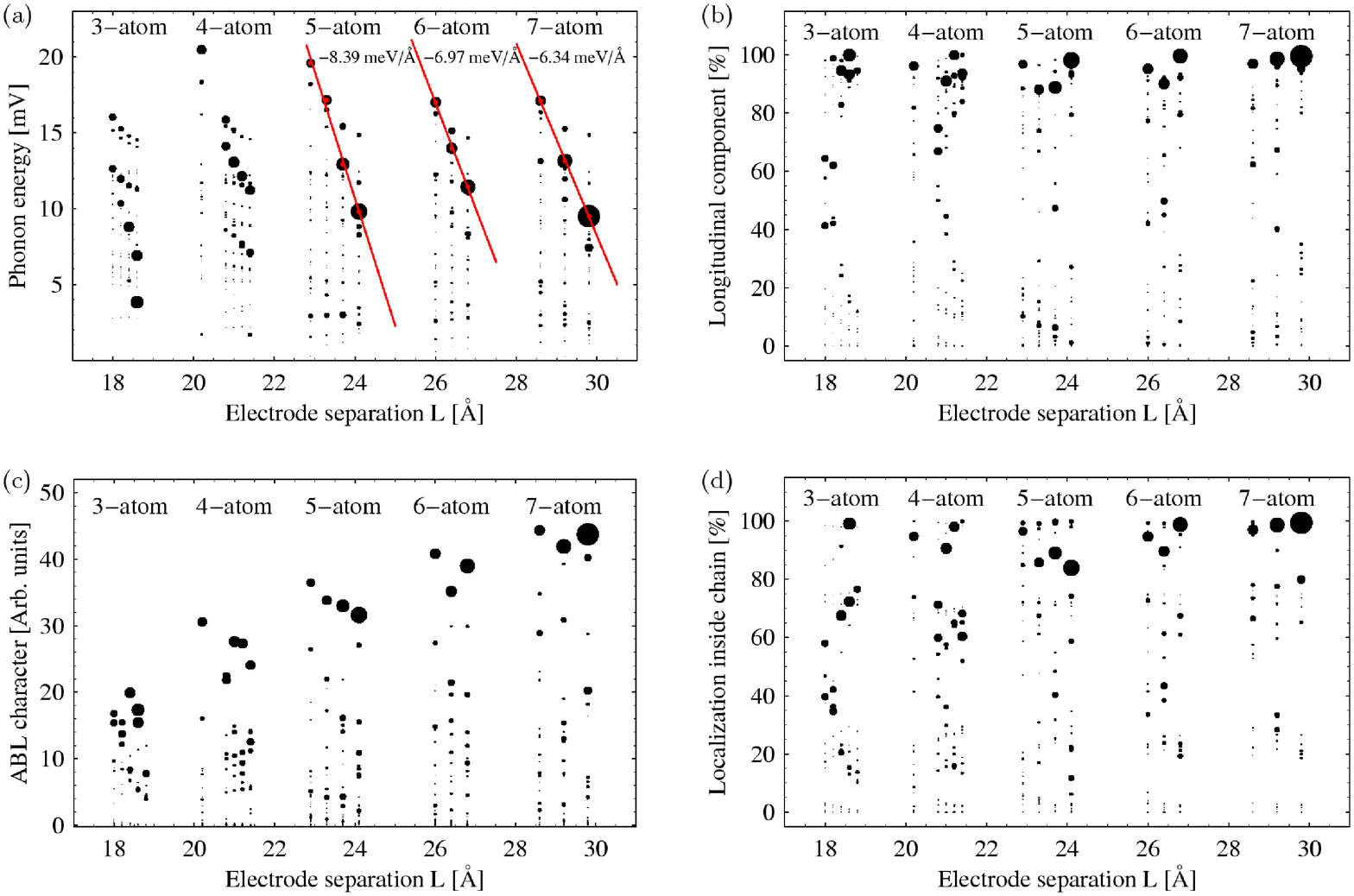}
  \caption{(Color online) Inelastic signals plotted as a function of the electrode separation
  $L$. Each mode is represented by a dot with an area proportional to
  the corresponding conductance drop. On the $y$-axis we show (a) the
  phonon mode energy, (b) a measure of the longitudinal component of the mode,
  (c) a measure of the ABL character, and (d) a measure of the
  localization to the wire atoms only. The straight lines in plot (a) are
  linear interpolations to the most significant signals
  (the slopes are given too).}
  \label{fig:InelasticSignalsVsL}
\end{figure*}

We start out by showing that the LOE and SCBA approaches essentially
predict the same inelastic signals for atomic gold wires, thereby
reducing the computational expense in the detailed analysis to
follow. For this purpose only we consider a computationally reduced
problem where the device and dynamic atoms regions are minimized as
compared with those generally adopted in this section. We will thus
simply allow the wire atoms to vibrate and take the device space as
the wire plus pyramidal bases only. Compared with the electronic
structure and phonon energies the thermal energy typically sets the
smallest energy scale for variations in the Green's functions etc.
Instead of using the experimentally relevant temperature of $T=4.2$
K (or even less) we further simplify the calculations by taking
$T=10.0$ K for the moment since this requires fewer points on the
energy grid, cf.~\Appref{sec:SCBAappendix}.

The differential conductances as resulting from evaluating
\Eqref{eq:MeirWingreenCurrent} with and without SCBA phonon
self-energies as well as evaluating the LOE expression
\Eqref{eq:current1} are shown in \Figref{fig:SCBAvsLOEforAuWires}.
The dotted curve is the purely elastic result (no phonon
self-energy) and the circles the full SCBA (including all
vibrational modes in the externally damped limit
$\gamma_\textrm{damp}\gg \gamma_\textrm{e-h}$ of
\Secref{sec:DampingRegimes}). The red line corresponds to the LOE.
The elastic conductance displays a slight variation with bias that
relates to the weak energy dependence in the zero-bias transmission
function at the $\Gamma$-point. The full SCBA calculation clearly
shows two symmetric conductance drops which are due to inelastic
scattering against vibrations (we will return later to a discussion
of the physics). The LOE calculation does not include the elastic
variation but gives basically the same predictions for the inelastic
signals. This is clear from a comparison with the SCBA where the
elastic background signal has been subtracted (dashed curve). Based
on a number of such tests, and the fact that the e-ph couplings are
weak (or more precisely, that the inelastic signal is a small change
in conductance of the order 1-2 \%), we conclude that the
approximations leading to the LOE expressions are valid in the case
of atomic gold wires. To appreciate this fact, we note that the SCBA
curves in \Figref{fig:SCBAvsLOEforAuWires} required approximately 40
CPU-hours in a parallel job running on 4 processors whereas the LOE
results only required a few seconds on one processor. The LOE
approach is thus justified for a full analysis of the 3- to 7-atom
gold wires.

Figure~\ref{fig:LOEcalculationsForAu} shows the calculated
differential conductance of the 3- to 7-atom wires under different
electrode separations $L$ and in the externally damped limit. The
device region and dynamic atoms are here as indicated in
\Figref{fig:GenericTransportSetup}, and the temperature of the leads
is $T=4.2$ K. The curves display symmetric drops at voltages
corresponding to particular phonon energies. The dominant inelastic
signal moves towards lower energies and increase in magnitude as the
wires are elongated. Furthermore, sometimes also a secondary feature
is found below 5 meV, e.g., Figs.~\ref{fig:SCBAvsLOEforAuWires} and
\ref{fig:LOEcalculationsForAu}. These observations are also
characteristic for the
experiments,\cite{AgUnRu.02.Onsetofenergy,AgUnRu.02.Electrontransportand}
and in agreement with previous
calculations.\cite{FrBrLo.04.InelasticScatteringand,ViCuPa.05.Electron-vibrationinteractionin}

To extract the general trends on how the inelastic signal depends on
details in the atomic arrangement we present in
\Figref{fig:InelasticSignalsVsL} our calculated data in different
forms. In these plots we represent each phonon mode by a dot with an
area proportional to the corresponding conductance drop. The
abscissa corresponds to the electrode separation whereas the
ordinate is used to highlight certain properties of the vibrational
modes. In this way, \Figref{fig:InelasticSignalsVsL}(a) illustrates
the mode frequency change with electrode separation. From a linear
fit to the strongest signals we predict a frequency shift of $-8.45$
meV/{\AA} for the 5-atom wire falling off to $-6.34$ meV/{\AA} for
the 7-atom long wire. Further, to understand the nature of the modes
that influence the electronic transport we can try to quantify some
important characteristics. As it has previously been shown,
longitudinal modes with an alternating bond length (ABL) character
are expected to be the dominating
ones.\cite{AgUnRu.02.Onsetofenergy,FrBrLo.04.InelasticScatteringand,
FrBrLo.04.ModelingofInelastic} To measure the longitudinal part of a
given vibrational mode $\mathbf{v}^\lambda$ we define a sum over
$z$-components $\sum_{I}(\textrm{v}^\lambda_{Iz})^2\leq 1$ where $I$
runs over all dynamic atoms (the upper bound is due to the
eigenmodes normalization
$\mathbf{v}^\lambda\cdot\mathbf{v}^\lambda=1$). This quantity is
shown in \Figref{fig:InelasticSignalsVsL}(b). The plot clearly
expresses that the modes with the largest signals (large dot area)
also have a strong longitudinal component. Further, to show that
these modes also have ABL character, we also define a sum $\sum_{I>
J} |\textrm{v}^\lambda_{Iz}-\textrm{v}^\lambda_{Jz}|$ where $I$ and
$J$ are nearest neighbor atoms in the chain. This second quantity is
shown in \Figref{fig:InelasticSignalsVsL}(c), from which we learn
that the important modes also have the largest ABL measure (the
absolute scale is irrelevant).

Another important aspect is whether the modes are really localized
in the wire or not. Remember that our approach assumes that atoms
outside the dynamic region are fixed. Therefore, if we have
eigenvectors with significant amplitude near the boundary of the
dynamic region, this assumption does not seem to be valid (most
likely the eigenvector is not a true eigenvector of the real
system). In other words, we want to make sure that the modes which
are responsible for the inelastic scattering are sufficiently
localized ``deep'' inside the dynamic region. To show this we
calculate $\sum_{I}\mathbf v^\lambda_I\cdot \mathbf v^\lambda_I\leq
1$ where $I$ runs over the 3 to 7 wire atoms. This quantity is
represented in \Figref{fig:InelasticSignalsVsL}(d) and confirms that
indeed the important modes are localized in the chain; particularly
for the 5-, 6-, and 7-atom wires the localization is almost perfect.

In conclusion, from the results presented in
\Figref{fig:InelasticSignalsVsL}, we learn that the inelastic signal
in the conductance is effectively described by a simple selection
rule in which longitudinal vibrational modes with ABL mode
character---localized in the wire---are the main cause of the
inelastic scattering. We are further able to quantify the frequency
down-shift and signal increase with strain.

\subsection{Vibrational lifetimes and local heating}
\label{sec:PhononLifetimeDiscussion}

\begin{figure}[t!]
  \centering
  %\begin{picture}(0,0)
  %  \put(-20,37){(a)}
  %  \put(-20,-25){(b)}
  %\end{picture}
  %\includegraphics[width=0.8\columnwidth]{figures/S29.20_ABLmode.eps}\vspace{5mm}
  %\includegraphics[width=1.\columnwidth]{figures/S29.20_ABLmodeBroadening.eps}
  \includegraphics[width=\columnwidth]{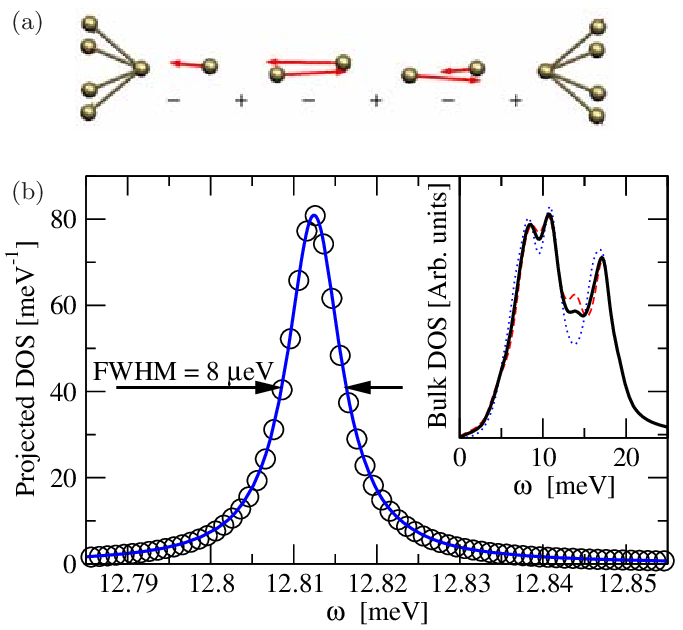}
  \caption{(Color online) ABL-mode broadening due to coupling to bulk phonons. The
  spectrum $B_\lambda(\varepsilon)$ corresponds to the important ABL-mode for
  a 7-atom wire ($L=29.20${\AA}).
  By fitting the calculated points with a Lorentzian we extract a full-width half
  max (FWHM) broadening of $2\gamma_\textrm{damp}^\lambda=8 \mu$eV
  and a frequency shift of $\delta\omega_\lambda=-6\mu$eV. The inset shows the calculated
  total density of states for bulk Au (full line), as well as a decomposition in
  the direction of the transport (dashed red curve) and in the transverse direction
  (dotted blue curve).}
  \label{fig:S29.20-ABLmode}
\end{figure}

From \Figref{fig:InelasticSignalsVsL}(d) we get a hint about the
damping of the modes from the coupling to bulk phonons. If a mode is
localized ``deep'' inside the dynamic region this coupling is
negligible and the mode is expected to have a long life-time, i.e.,
to be weakly damped by the coupling to the bulk. As discussed in
\Secref{sec:introPhononDOSprojection} we can estimate this damping
from the width of the phonon density of states projected onto the
mode vector.

As an illustration of this approach, we calculate the damping of the
dominating ABL mode according to \Eqref{eq:LorentzBroadening} in the
case of the 7-atom wire with electrode separation $L=29.20$ {\AA}.
This mode, shown in \Figref{fig:S29.20-ABLmode}(a), has a
localization quantity (as defined above) of value 0.987, i.e., it is
98.7{\%} localized in the wire. We begin by determining the dynamic
matrix of the whole wire supercell [\Figref{fig:Structures}(e)] as
described in \Secref{sec:VibModes}. To describe the bulk properties
of gold we pick the intra-layer and inter-layer elements (inside the
slab) in the dynamic matrix along the transport direction, and use
recursive techniques to calculate bulk and surface phonon Green's
functions. Because of periodicity in the transverse plane---which
gives rise to artificial sharp resonances in the spectrum---we
broaden the phonon Green's functions by taking $\eta=1.0$ meV. This
approach leads to the total phonon density of states (full black
line) shown in the inset of \Figref{fig:S29.20-ABLmode}. This shape
compares reasonably well with other calculations and
experiments.\cite{LYSMNI.73.LATTICE-DYNAMICSOFGOLD,TRDE.85.BULKANDSURFACE}
The inset also shows the phonon density of states decomposed in the
direction of the transport (dashed red curve) as well as in the
transverse directions (dotted blue curve); the observed isotropy
that is expected for bulk is actually quite satisfactory. Finally,
we calculate the projected phonon density of states
$B_\lambda(\omega)$ for the ABL mode of interest according to
\Eqref{eq:LorentzBroadening}. This projection on a discrete energy
grid is shown in \Figref{fig:S29.20-ABLmode} (open circles). By
fitting a Lorentzian to the calculated data points we obtain a FWHM
of $8~\mu$eV and a shift in frequency by $-6~\mu$eV. Based on these
calculations we thus estimate the phonon damping to be of the order
$\hbar\gamma_\text{damp}^\lambda = 4~\mu$eV. In fact, this is rather
a lower bound, since we have not included anharmonic contributions
etc.\cite{Mi.06.Anharmonicphononflow} However, compared with the
phonon energy we see that indeed $\gamma_\text{damp}^\lambda \ll
\omega_\lambda$, and thus that the use of free phonon Green's
functions in the SCBA self-energy \Eqref{eq:SigmaLessGreat} is
justified.

\begin{figure}[t!]
  \centering
  \includegraphics[width=\columnwidth]{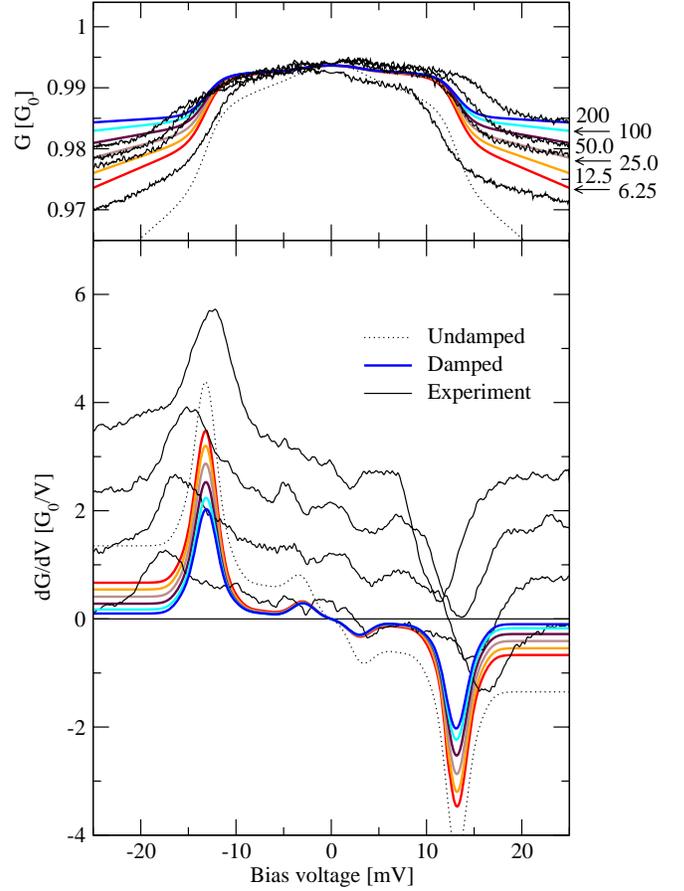}
  \caption{(Color online) Comparison between theory and experiment
  (Ref.~\onlinecite{AgUnRu.02.Electrontransportand}) for the inelastic conductance of
  an atomic gold wire. The measured characteristics (noisy black curves) correspond
  to different states of strain of wire (around 7 atoms long). The calculated
  results (smooth colored lines) are for the 7-atom wire at $L=29.20$ {\AA}
  using different values for the external damping as indicated in
  the upper right corner of the plot (in units of
  ${\mu}$eV$/\hbar$). The dashed curve is the calculated result in
  the externally undamped limit ($\gamma_\textrm{damp}^\lambda=0$). The lower plot
  is the numerical
  derivative of the conductance, where the experimental curves have
  been offset by multiples of G$_0$/V for clarity. Note the indication of a
  secondary phonon feature below 5 meV in all curves. The temperature is $T=4.2$ K and
  the lock-in modulation voltage $V_\textrm{rms}=1$ meV (in both theory and experiment).}
  \label{fig:TheoryVsExperiment}
\end{figure}

Let us next investigate the implications of a finite phonon lifetime
on the local heating. This is done by solving the rate equation
\Eqref{eq:RateEquation} for the mode occupation at a fixed bias
voltage. For instance, the inelastic conductance characteristics
(including heating) for our 7-atom wire are shown in
\Figref{fig:TheoryVsExperiment} for different values of the phonon
damping $\gamma_\text{damp}^\lambda$ (smooth colored lines). As seen
in the figure, and as we have shown
previously,\cite{FrBrLo.04.InelasticScatteringand} the effect of the
heating is to introduce a slope in the conductance beyond the phonon
threshold voltage. This is because the nonequilibrium mode
occupation increases the number of scattering events of the
traversing electrons. Consequently the conductance goes down as the
bias (and hence the occupation level) increases. The smaller the
damping, the more the mode occupation is driven out of equilibrium,
i.e., to a larger average excitation level. In the extreme case of
no damping $\gamma_\text{damp}^\lambda=0$ (dotted curve) [the
externally undamped limit in
Ref.~\onlinecite{FrBrLo.04.InelasticScatteringand}], the local
heating is maximal. On the other hand, a sufficiently large damping
may effectively prevent phonon heating [the externally damped limit
in Ref.~\onlinecite{FrBrLo.04.InelasticScatteringand}]. From
\Figref{fig:TheoryVsExperiment} we see that with a phonon damping as
large as 200 $\mu$eV/$\hbar$ the slope has vanished.

Figure~\ref{fig:TheoryVsExperiment} also compares our theoretical
results to the original experimental measurements by Agra\"it
\etal\cite{AgUnRu.02.Onsetofenergy} (noisy curves). The four
experimental characteristics (aligned with the calculated zero-bias
conductance) corresponds to a presumably 7-atom long gold wire under
different states of strain recorded at low temperatures $T=4.2$ K.
From this plot it is clear that theory and experiment are in
excellent agreement with respect to the position of the phonon
signal and the magnitude of the dominant drop. One also notices the
indication of a secondary phonon feature below 5 meV in all curves.
But what is particularly interesting is that the measured
conductance slopes beyond the threshold seem to agree well with a
phonon damping of the order 5-50 $\mu$eV, which is further quite
reasonable according to our estimate above. The only feature which
is not perfectly reproduced is the experimental width of phonon
signal lineshape---as seen from the derivative of the conductance
$dG/dV$ in the lower part of the figure---which is somewhat wider
than the calculated ones (which for comparison also includes the
instrumental lock-in broadening corresponding $V_\textrm{rms}=1$
meV).

%%%%%%%%%%%%%%%%%%%%%%%%%%%%%%%%%%%%%%%%%%%%%%%%%%%%%%%%%%%%%%%%%%%%%%%%%%%%%%%%%%%%%
%\input{HydrocarbonMolecules}
%%%%%%%%%%%%%%%%%%%%%%%%%%%%%%%%%%%%%%%%%%%%%%%%%%%%%%%%%%%%%%%%%%%%%%%%%%%%%%%%%%%%%

%%%%%%%%%%%%%%%%%%%%%%%%%%%%%%%%%%%%%%%%%%%%%%%%%%%%%%%%%%%%%%%%%%%%%%%%%%%%%%
\section{Hydrocarbon molecules between gold contacts}
\label{sec:HydrocarbonMolecules} The general method described in
\Secref{sec:ElectronicStructureMethods} and \ref{sec:Transport} is
applicable to many other systems than atomic gold wires. Examples of
systems where it is interesting to apply this method include wires
and contacts of other metals as well as individual molecules. In
fact, we have already used the present method to study conjugated
and saturated hydrocarbon molecules in between gold surfaces, see
Ref.~\onlinecite{PaFrBr.06.InelasticTransportthrough}. The purpose
of this section is to illustrate that our method is general enough
to apply to many systems; especially that the LOE approximation is
likely to be valid for a range of systems where, at first glance, it
is not expected to work.

We start with a brief description of our previous
results\cite{PaFrBr.06.InelasticTransportthrough} motivated by the
recent experiments by Kushmerick
\etal\cite{KuLaPa.04.Vibroniccontributionsto} They measured the
inelastic scattering signal through three different molecules (C11,
OPV, and OPE) connected to gold electrodes by means of a cryogenic
crossed-wire tunnel junction setup. Since the number of molecules
present in the experimentally realized junctions is unknown it is
advantageous to look at the inelastic electron tunneling
spectroscopy (IETS) signal defined as
\begin{equation}
\mathrm{IETS} \equiv \frac{{\intd}^2 I/{\intd}V^2}{{\intd}
I/{\intd}V},
\end{equation}
which---if the current $I$ simply scales with the number of
molecules---is independent of the number of molecules in the
junction.

\begin{figure}[t!]
  \centering
  \includegraphics[width=.8\columnwidth]{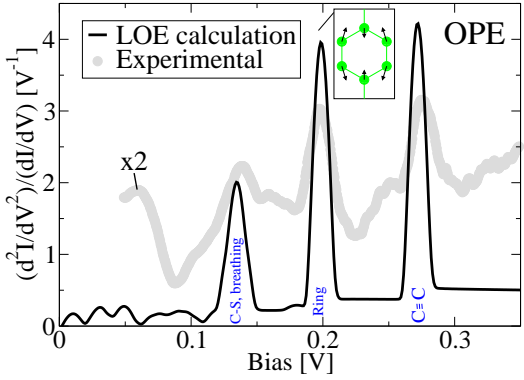}
  \caption{(Color online) Calculated IETS spectrum for an OPE molecule compared to the
  experimental data from
    Ref.~\onlinecite{KuLaPa.04.Vibroniccontributionsto}. Each of the three inelastic scattering
    peaks arise from different kinds of vibrations localized on the molecule.}
  \label{fig:OPE-LOE}
\end{figure}

\begin{figure}[t!]
  \centering
  %\begin{picture}(0,0)
  %  \put(-10,110){(\textsf{a})}
  %  \put(-10,-20){(\textsf{b})}
  %\end{picture}
  %\includegraphics[width=.8\columnwidth]{figures/OPVfig.eps}
  %\includegraphics[width=.8\columnwidth]{figures/OPEfig.eps}
  \includegraphics[width=\columnwidth]{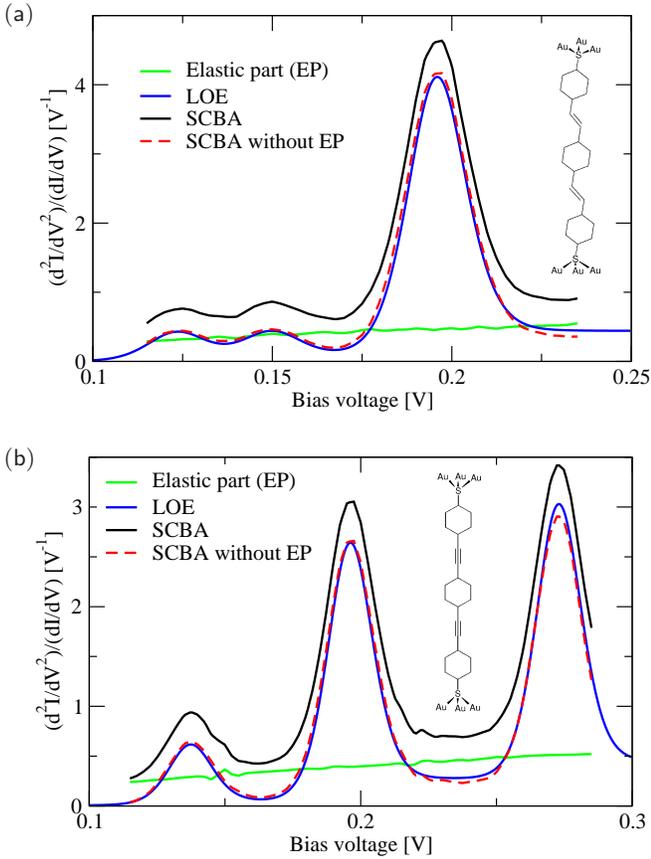}
  \caption{(Color online) Calculated IETS spectra for (a) an OPV molecule and (b) an OPE molecule. The
  chemical structure of these hydrocarbon molecules are shown in the insets. The two plots show that
  the simple LOE scheme predicts the same IETS spectrum as the full SCBA (if one neglects
  the elastic variation).}
  \label{fig:OPE-OPV}
\end{figure}

In Ref.~\onlinecite{PaFrBr.06.InelasticTransportthrough}, we used
the present LOE method to model the IETS spectra for each of these
three molecules. As an example, \Figref{fig:OPE-LOE} shows the
calculated and measured IETS spectrum in the case of the conjugated
OPE molecule [inset of \Figref{fig:OPE-OPV}(b)]. It is seen that our
theory reproduces the positions and relative heights of the
inelastic scattering peaks. The three main peaks are given by four
types of vibrations; one type is affecting the C--S stretch whereas
the other three involve the distortion of the C backbone of the
molecule. In our calculation the region of dynamic atoms includes 54
atoms corresponding to 162 vibrational modes (18 Au surface atoms
and 36 atoms in the molecule). We thus see that the IETS spectrum
must be related to certain selection rules that describe why only a
few vibrational modes affect the current. These selection rules may
be understood from studying the electron scattering states and the
symmetry of the e-ph interaction.\cite{Paulsson} For the other two
molecules (OPV and C11) we found a similar good agreement with the
experiments by Kushmerick {\etal} However, the transmission
$T(\varepsilon)$ through these three molecules is actually varying
significantly with energy, since the electron conduction process
involves states around the Fermi energy that lie in the gap between
the molecular levels. For instance, in an energy window of 0.4 eV
this variation is of the order
$T(\varepsilon_\text{F}-0.2\textrm{eV})/T(\varepsilon_\text{F}+0.2\textrm{eV})\approx
4$ for the OPE molecule. Accordingly the use of the LOE
approximation might seem inappropriate for these systems. With a
detailed comparison between LOE and full SCBA calculations
(including this energy dependence) we can nevertheless show that the
LOE approximation provides effectively the same prediction for the
IETS spectrum. This comparison is found in \Figref{fig:OPE-OPV}.

Since the SCBA is computationally expensive it is not realistic to
use the same high accuracy as for LOE calculations. We therefore
reduce the device subspace and the region of dynamic atoms to
include only the molecule. Furthermore we use a smaller SZP basis
set describing the OPE (OPV) molecule reducing the device subspace
to 264 (280) atomic orbitals. Finally we include only the 5 (3) most
important vibrational modes (selected from a LOE calculation). With
these simplifications we calculated the current for 81 (61) bias
points using an average of 9 (8) iterations to converge the SCBA on
an energy grid of approximately 500 points. These SCBA calculations
required 40 (18) hours on 10 Pentium-4 processors working in
parallel. For comparison, the corresponding LOE calculations can be
performed in less than 1 minute on a single Pentium-4 processor.

The results shown in \Figref{fig:OPE-OPV} reveal that the LOE
approximation captures the inelastic scattering signal with a very
satisfactory accuracy. The main discrepancy between LOE and SCBA is
directly related to the elastic part of the transport which can
easily be corrected for without solving the full SCBA equations,
cf.~\Secref{sec:INelastictransmission}. We have thus used our
implementation of SCBA to justify that the simpler LOE scheme can
actually be applied for the IETS spectra of the hydrocarbon
molecules. This is not a trivial result because the energy variation
in the transmission around the Fermi energy for these systems seems
to violate one of the fundamental assumptions of the LOE.

%%%%%%%%%%%%%%%%%%%%%%%%%%%%%%%%%%%%%%%%%%%%%%%%%%%%%%%%%%%%%%%%%%%%%%%%%%%%%%%%%%%%%
%\input{Conclusions}
%%%%%%%%%%%%%%%%%%%%%%%%%%%%%%%%%%%%%%%%%%%%%%%%%%%%%%%%%%%%%%%%%%%%%%%%%%%%%%%%%%%%%
%%%%%%%%%%%%%%%%%%%%%%%%%%%%%%%%%%%%%%%%%%%%%%%%%%%%%%%%%%%%%%%%%%%%%%%%%%%%%%
\section{Conclusions and outlook}
\label{sec:Conclusions} In this paper we have presented a
first-principles method for calculating the effects of vibrations
and e-ph couplings in the electronic transport properties of an
atomic-scale device. Our implementation that extends the {\siesta}
implementation of Kohn-Sham DFT and the {\transiesta} scheme for
elastic transport is described in detail, highlighting the important
computational steps for the complete analysis. The inelastic
transport problem is addressed using the NEGF formalism with the
e-ph interaction treated up to the level of SCBA. We also describe
the computationally simple LOE scheme. As illustrations of the
methodology we have applied it to model the phonon signals in the
conductance of atomic gold wires and hydrocarbon molecules between
gold surfaces. In both cases the comparison with experimental
results is very satisfactory. While we expect our method to be
successful for a wide range of nanoscale systems, there are also
some important aspects where further research and development may
lead to improvements. We therefore close this paper with an outlook
of some of the challenges we believe are important.

%%% outlook
While we have argued that the vibrations for the systems considered
here are reasonably well described by free phonon Green's functions,
there might also be situations where the phonon system has to be
treated beyond free dynamics, e.g., by including self-energies from
e-h pair damping, anharmonic phonon-phonon couplings (inside the
device), and resonant phonon-phonon couplings (between device and
electrodes). And as we have shown in this work, these precise
damping conditions of the phonons are governing the device heating.
Another issue is the bias-induced changes in geometry and e-ph
couplings. Further development along these lines might thus lead to
a better understanding of transport in the high-bias regime. On the
more technical side, it would be interesting to extend the present
scheme to describe the interplay between e-ph couplings and other
delicate effects such as spin-polarized currents, spin-orbit
couplings, etc. For instance, phonon heating could mediate an
important effective interaction between the two spin channels.

In conclusion, the present paper contributes to the evolving
understanding of phonon scattering and local heating in nanoscale
systems. These effects are important to elucidate the structural
properties from the electronic transport characteristics and
ultimately for the stability of devices.

%%%%%%%%%%%%%%%%%%%%%%%%%%%%%%%%%%%%%%%%%%%%%%%%%%%%%%%%%%%%%%%%%%%%%%%%%%%%%%
\section*{Acknowledgments}
During the development of our scheme many people contributed
directly or indirectly to our work. In particular, we are grateful
to N.~Lorente for many invaluable discussions, and thank
J.~C.~Cuevas, H.~Ness and T.~Todorov for useful comments. The
authors are thankful to N.~Agra\"it, D.~Djukic, and J.~M.~van
Ruitenbeek for many stimulating discussions on their respective
experiments of phonon scattering in atomic-scale contacts.
M.P.~would like to thank S.~Datta for his insight into heating
aspects in quantum transport. T.F.~thanks T.~Novotn\'{y} for
guidance in our early numerical implementation of the Hilbert
transform.

This work, as a part of the European Science Foundation EUROCORES
Programme SASMEC, was partially supported by funds from the SNF and
the EC 6th Framework Programme. Computational resources were
provided by the Danish Center for Scientific Computing (DCSC).

%%%%%%%%%%%%%%%%%%%%%%%%%%%%%%%%%%%%%%%%%%%%%%%%%%%%%%%%%%%%%%%%%%%%%%%%%%%%%%%%%%%%%
%\input{Appendix}
%%%%%%%%%%%%%%%%%%%%%%%%%%%%%%%%%%%%%%%%%%%%%%%%%%%%%%%%%%%%%%%%%%%%%%%%%%%%%%%%%%%%%
%%%%%%%%%%%%%%%%%%%%%%%%%%%%%%%%%%%%%%%%%%%%%%%%%%%%%%%%%%%%%%%%%%%%%%%%%%%%%%
\appendix

\section{Hilbert Transform}
\label{sec:Hilbert} The purpose of this appendix is to discuss
efficient numerical ways to approximate the Hilbert transform of a
continuous function $f(x)$, here defined as\footnote{We hereby
correct a sign error in our definition of the Hilbert transform in
Ref.~\onlinecite{PaFrBr.05.Modelinginelasticphonona}.}
\begin{eqnarray}
\mathcal {H}_x\{f\}(y)=\frac 1\pi
\mathcal{P}\int_{-\infty}^\infty\intd x \frac{f(x)}{x-y},
\label{eq:90:defHilbert}
\end{eqnarray}
where $\mathcal P$ denotes the Cauchy principal value integral.

We approximate the function $f(x)$ by a linear interpolation
$f_I(x)$ to the values $f_i=f(x_i)$ known at the discrete grid
points $\{x_i\}$. This we can write in the following way
\begin{eqnarray}
f(x) \approx f_I(x) &\equiv&\sum_{i=1}^N f_i\psi_i(x),
\end{eqnarray}
where the kernel function associated with the linear interpolation
is
\begin{eqnarray}
\psi_i(x) &=&
\frac{x-x_{i-1}}{x_i-x_{i-1}}[\theta(x_i-x)-\theta(x_{i-1}-x)]\nonumber\\
&&+\frac{x_{i+1}-x}{x_{i+1}-x_{i}}[\theta(x_{i+1}-x)-\theta(x_{i}-x)].\qquad
\end{eqnarray}
On this form we implicitly assume that the function falls off to
zero at the ends of the grid, i.e., that the function has finite
support. We can then approximate the Hilbert transform of $f(x)$ by
the Hilbert transform of $f_I(x)$, i.e.,
\begin{eqnarray}
\mathcal{H}_x\{f\}(x_j)&\approx& \mathcal{H}_x\{f_I\}(x_j)\nonumber\\
&=&\frac 1\pi \mathcal{P}\int_{-\infty}^\infty\intd x \frac{f_I(x)}{x-x_j}\nonumber\\
&=& \sum_{i=1}^N  K_{ji} f_i,
\end{eqnarray}
where we have identified a transformation kernel
\begin{eqnarray}
K_{ji} &\equiv & \frac 1\pi \mathcal{P}\int_{-\infty}^\infty\intd
x \frac{\psi_i(x)}{x-x_j}\nonumber\\
&=& \frac 1\pi \Big[
\frac{x_j-x_{i-1}}{x_i-x_{i-1}}\ln{\Big(\frac{x_i-x_{j}}{x_{i-1}-x_j}\Big)}\nonumber\\
&&\quad+\frac{x_{i+1}-x_{j}}{x_{i+1}-x_{i}}\ln{\Big(\frac{x_{i+1}-x_{j}}{x_{i}-x_j}\Big)}\Big].
\end{eqnarray}
Having determined the matrix $K_{ji}$ corresponding to a given grid
$\{x_i\}$, the Hilbert transform amounts to a matrix-vector product
operation. With $N$ grid points this scales as $\mathcal O (N^2)$.

A typical situation is that of an equidistant grid
$x_i-x_{i-1}=\Delta$ (for all $i$), where a more effective algorithm
can be devised. In this case we can write $x_i-x_{j}=(i-j) \Delta$,
and the kernel function, that becomes a function of the index
difference $m=j-i$ only, reduces to
%\begin{eqnarray}
%k_{j-i} &=& \frac 1\pi \Big[
%(j-i+1)\ln{\Big(\frac{i-j}{i-1-j}\Big)}\nonumber\\
%&&\quad+(i+1-j)\ln{\Big(\frac{i+1-j}{i-j}\Big)}\Big].
%\end{eqnarray}
\begin{eqnarray}
K^\Delta_{m} &=& \frac 1\pi\big[-(m-1)\ln(m-1)\nonumber\\
&& \quad +2m\ln m-(m+1)\ln(m+1)\big].
\end{eqnarray}
The Hilbert transform $\mathcal{H}_x\{f_I\}(x_j)= \sum_{i=1}^N
K^\Delta_{j-i} f_i$ has then taken the form of a discrete
convolution which effectively can be calculated with the Fast
Fourier transform (FFT) algorithm. This scales only as $\mathcal O
(N\ln N)$.

%%%%%%%%%%%%%%%%%%%%%%%%%%%%%%%%%%%%%%%%%%%%%%%%%%%%%%%%%%%%%%%%%%%%%%%%%%%%%%%%%%%%%
\section{Numerical implementation of SCBA}
\label{sec:SCBAappendix} Calculating the current numerically using
the SCBA is highly nontrivial for large systems. This appendix
discusses our solutions to the main difficulties encountered within
the SCBA. We exemplify the size and scope of the calculations, e.g.,
the sizes of matrices and the energy grid, with values taken from
the SCBA calculation presented in \Secref{sec:HydrocarbonMolecules}
on the OPE molecule.

The current and power expressions \Eqref{eq:MeirWingreenCurrent} and
(\ref{eq:power}) are integrated numerically using a third order
polynomial interpolation. Since the inelastic signal is typically
small, the current has to be determined with a high accuracy, which
implies a fine resolution of the energy grid for the integration.
Further, the range of this grid has to include not only the bias
window but also additional energies due to the nonlocal character
(in energy) of the Hilbert transform, cf.~\Eqref{eq:SigmaRet}. These
limitations make a nonuniform grid preferable. We thus construct a
dense grid around each of the important energies
$\varepsilon=\mu_{L,R},\, \mu_{L,R} \pm \hbar \omega_\lambda,\ldots$
and a coarser one elsewhere. The resolution of the fine grid is
determined by the temperature and should have a point separation
around $\delta\varepsilon \leq 0.5~\text{k}_\text{B}T$. For the OPE
molecule we found it adequate at $T=40$ K to use a fine grid with
$\delta\varepsilon =1.7$ meV and a coarse grid with
$\Delta\varepsilon=10.0$ meV spanning the energy range $[-0.5,0.5]$
eV. With a nonuniform grid the necessary number of energy points may
thus be reduced.

The solution of the SCBA approximation requires substantial amounts
of CPU time and memory. Analyzing the memory requirements we find
that we need to retain $\mathbf G^{\lessgtr,r}(\varepsilon)$ and
$\mathbf\Sigma^{\lessgtr,r}_\mathrm{ph}(\varepsilon)$ in memory.
Each of these matrices requires a memory allocation of $\mathcal O
(N_\textrm{grid}\,N_\textrm{basis}^2)$ bytes, where
$N_\textrm{grid}$ is the number of grid points, and
$N_\textrm{basis}$ the size of the electronic basis. For the OPE
calculation in \Secref{sec:HydrocarbonMolecules} each matrix takes
up 500 Megabytes of memory (500 energy points $\times$ $250^2$
matrix size $\times$ 16 bytes/complex number). In addition to the
demanding memory requirement, significant computational time (400
CPU hours in total) is needed.

The computationally heaviest part is the calculation of
\Eqref{eq:SigmaLessGreat}, which we rewrite as
\begin{eqnarray}
\label{eq.mpnPhSig}
\mathbf{\Sigma}^{\lessgtr}_\mathrm{ph}(\varepsilon)&=& \sum_\lambda
\mathbf{M}_\lambda \Big[ \langle n_\lambda \rangle
\mathbf{G}^\lessgtr (\varepsilon \pm \hbar
\omega_\lambda) \\
&& \qquad\qquad + (\langle n_\lambda\rangle + 1) \mathbf{G}^\lessgtr
(\varepsilon\mp \hbar \omega_\lambda) \Big]
\mathbf{M}_\lambda.\nonumber
\end{eqnarray}
From this equation we see that the CPU time scales as
$\mathcal{O}(N_\textrm{ph}\,N_\textrm{grid}\,N_\textrm{basis}^3\,N_\textrm{iter})$
[since each matrix multiplication scales as
$\mathcal{O}(\,N_\textrm{basis}^3)$], where $N_\textrm{ph}$ is the
number of vibrational modes and $N_\textrm{iter}$ the number of
iterations needed for self-consistency of the SCBA.

We have overcome the memory and computational requirements by a
parallelization of our computer code by dividing the energy grid
over the available processors. The only significant complication is
the evaluation of \Eqref{eq.mpnPhSig}, where quantities couple
across the energy division. To overcome this, we first redistribute
the Green's functions $\mathbf G^\lessgtr(\varepsilon)$ over the
processors by changing from energy division to matrix indices
division. Then the energy-shifted Green's functions can be added for
each matrix index. Next we transform the outcome back to energy
division and carry out the matrix multiplications with $\mathbf
M_\lambda$. We have implemented this procedure efficiently in a way
that lets the necessary communication occur while other calculations
are running, i.e., while the lesser part of the equation is being
communicated between processors, the matrix multiplications for the
greater part are being computed and vice versa. In practice, this
parallelization works very well and the computation time scales
almost linearly with the number of processors.

%%%%%%%%%%%%%%%%%%%%%%%%%%%%%%%%%%%%%%%%%%%%%%%%%%%%%%%%%%%%%%%%%%%%%%%%%%%%%%%%%%%%%
\section{Signal broadening by lock-in modulation voltage}
\label{sec:ddIlockin} As discussed in
\Secref{sec:broadening-mechanisms} the lock-in technique for
measuring the differential conductance (and derivatives) introduces
a broadening of the intrinsic current-voltage characteristics due to
a finite modulation voltage. The basic idea is to measure the
frequency components of the current at multiples of the applied
harmonic modulation, since these relates to the derivatives of the
current. Following Hansma,\cite{HA.77.INELASTICELECTRON-TUNNELING}
we can analytically write the frequency components as the following
averages over an oscillation period
\begin{eqnarray}
I_{\omega} &\equiv& \frac{\omega}{\pi A} \int\limits_0^{2\pi
/\omega}
    I[V+A\cos(\omega t)] \cos(\omega t) \, \intd t   \nonumber       \\
&=& \frac{2}{\pi} \int\limits_{-1}^{1}
     \frac{{\intd}I\left(V+A x \right)}{{\intd}V} \, \sqrt{1-x^2}
  \, {\intd}x,
\label{eq.dIlockin}
\end{eqnarray}
and
\begin{eqnarray}
I_{2\omega} &\equiv& \frac{4 \omega}{\pi A^2} \int\limits_0^{2 \pi
/\omega} I[V+A\cos(\omega t)] \cos(2 \omega t) \, {\intd}t   \nonumber       \\
&=& \frac{8}{3 \pi} \int\limits_{-1}^{1}
     \frac{{\intd}^2I \left(V+A x \right)}{{\intd}V^2} \, \left(1-x^2\right)^{3/2}
  \, {\intd}x,
\end{eqnarray}
where the modulation amplitude is $A=\sqrt{2}V_\textrm{rms}$.  The
partial integrations carried out above show that the components
$I_\omega$ and $I_{2\omega}$ are convolutions of the exact first and
second derivatives of the current with certain functions
proportional to $\sqrt{1-x^2}$ and $(1-x^2)^{3/2}$, respectively. If
we assume that the inelastic signal has no intrinsic width, the
inelastic conductance change is proportional to a step function
$\theta(eV-\hbar\omega_\lambda)$ and the second derivative to a
delta function $\delta(eV-\hbar\omega_\lambda)$. With these
functional forms the integrals can be evaluated, leading to a
modulation broadening of the first (second) derivative of
approximately 2.45 $V_\textrm{rms}$ (1.72 $V_\textrm{rms})$.

%%%%%%%%%%%%%%%%%%%%%%%%%%%%%%%%%%%%%%%%%%%%%%%%%%%%%%%%%%%%%%%%%%%%%%%%%%%%%%%%%%%%%
% Bibliography
%%%%%%%%%%%%%%%%%%%%%%%%%%%%%%%%%%%%%%%%%%%%%%%%%%%%%%%%%%%%%%%%%%%%%%%%%%%%%%%%%%%%%
%\bibliographystyle{apsrev}
%\bibliography{Bibliography}
%\input{ThisBib}

\end{document}